\documentclass[preprint2]{aastex}

\newcommand{\lsim}{\raisebox{-5pt}{$\;\stackrel{\textstyle<}{\sim}\;$}}

\slugcomment{To appear in Ap. J.  }

\lefthead{Vladilo \& al.}
\righthead{Ionization properties and   abundances  in DLAs }

\begin{document}

\title{ IONIZATION PROPERTIES AND ELEMENTAL ABUNDANCES  
IN DAMPED Ly\,$\alpha$ SYSTEMS}

\author{Giovanni Vladilo,  Miriam Centuri\'on, and Piercarlo Bonifacio,}
\affil{Osservatorio Astronomico di Trieste, Via G.B. Tiepolo 11, 34131 Trieste, Italy}

\author{J. Christopher Howk }
\affil{ Department of Physics and Astronomy, The Johns Hopkins University, 3400 North Charles Street,
Baltimore, MD 21218}


\begin{abstract}

We analyze extant data of  Al$^{2+}$, Al$^{+}$ and other low ions
with the aim of studying the
ionization properties of Damped Ly\,$\alpha$ systems (DLAs)
from the analysis of the    ratio 
$\cal R$(Al$^{2+}$/Al$^{+}$) $\equiv$ $N$(Al$^{2+}$)/$N$(Al$^{+}$).
We find good correlations  $\log N({\rm Al}^+)$ -  $\log N({\rm Si}^+)$
and $\log N({\rm Al}^+)$ -  $\log N({\rm Fe}^+)$ that we use   
to indirectly estimate $N$(Al$^{+}$)  from
$N$(Si$^{+}$) and/or $N$(Fe$^{+}$) measurements.  In this way we  
determine the   ratio 
$\cal R$(Al$^{2+}$/Al$^{+}$)  
for a sample of 20 DLAs.
Contrary to common belief, the ratio can attain relatively high values (up to 0.6), 
suggesting that gas of intermediate ionization state plays an important role in DLAs.  
On the other hand, the lack of any trend between 
abundance ratios, such as Si/H and Si/Fe, and $\cal R$(Al$^{2+}$/Al$^{+}$)
indicates that abundances are not severely influenced by ionization effects.   
We  find a     
$\log {\cal R}$(Al$^{2+}$/Al$^{+}$) - $\log N$(H$^{\circ}$) anticorrelation
that we use, in conjunction with idealized photoionization equilibrium calculations, 
to constrain the ionization properties and to predict ionization corrections
in DLAs. 
We consider two possible origins for the 
species of low and intermediate ionization  state:
(1)  neutral  regions devoid of Al$^{2+}$ and/or 
(2)  partially ionized, Al$^{2+}$-bearing regions.  
The $\log {\cal R}$(Al$^{2+}$/Al$^{+}$) - $\log N$(H$^{\circ}$) anticorrelation
can be naturally explained in terms of a two-region model with a soft, stellar-type
ionizing radiation field.  
We present abundance ionization corrections 
for 14 elements of astrophysical interest derived
with different types of ionizing spectra. 
For most of these   elements    the
corrections are generally below measurements errors,
contrary to the predictions of recent models presented in the literature.
We briefly discuss the potential effects of inaccuracies of the Al
recombination rates used in the photoionization calculations.

\end{abstract}


\section{Introduction}

The QSO absorption systems with hydrogen column densities
$N$(H$^{\circ}$) $> 2 \times 10^{20} $ cm$^{-2}$
--- called damped Ly$\alpha$ systems (DLAs) --- are believed
to originate in intervening galaxies or protogalaxies located at cosmological
distances (Wolfe et al. 1995). 
High resolution spectroscopy of  DLAs
is   a fundamental tool for probing the chemical and physical properties
of the associated high-$z$ galaxies and, more   generally, of the high-redshift
universe. 
In particular,   
abundance studies  have been performed in order
to probe the nature of DLA galaxies and
to cast light on the early stages of galactic evolution  
(Lu et al. 1996, Pettini et al. 1997, Prochaska \& Wolfe 1999, Molaro et al. 2000). 
In abundance studies it is important to take into account both
dust depletion and ionization  effects  
before deriving conclusions on the  nucleosynthetic processes at work
in DLA galaxies. 
The effects of dust depletion  have been investigated by several
research groups
(Lauroesch et al. 1996; Kulkarni, Fall \& Truran 1997;
Vladilo 1998, hereafter Paper I; Savaglio, Panagia \& Stiavelli 2000). 
All these studies indicate that
dust corrections can have a significant impact on our understanding
of the chemical history of DLAs. 
In order to cast light on the intrinsic abundance trends some authors have analyzed
DLAs with modest or negligible dust content (Pettini et al. 2000, Molaro et al. 2000)
or elements which are essentially unaffected by dust depletions (Centuri\'on et al. 1998).

In the present work we investigate the effects of ionization on the observed abundances. 
These effects are generally neglected 
because  the abundances are derived from  dominant ionization
states of the elements of interest. It is well known from Galactic interstellar studies
that 
the dominant ionization state in   H$^{\circ}$ regions is
the neutral one for elements with first IP $>$ 13.6 eV 
(e.g. O$^\circ$, N$^\circ$)
and the singly ionized one for elements with
first IP $<$ 13.6 eV and second IP $>$ 13.6 eV
(e.g. C$^+$, Mg$^+$, Al$^{+}$, Si$^+$, S$^+$, Cr$^+$, Mn$^+$, Fe$^+$, and Zn$^+$). 
The reason for this is that the bulk of the H$^{\circ}$ gas   is  self-shielded from  
h$\nu$ $>$ 13.6 eV photons,  but transparent to 
h$\nu$ $<$ 13.6 eV photons. 
The first studies of ionization balance in DLAs  
have   indicated that ionization corrections
are  negligible for the low ions used in abundance
determinations (Viegas 1995, Lu et al. 1995, Prochaska \& Wolfe 1996).
However, the presence of Al$^{2+}$ in DLAs suggests   
that ionization corrections may  be important. 
From the aluminium ionization potentials ---
IP(Al$^{\circ}$)=5.99 eV and  IP(Al$^{+}$)=18.83 eV --- 
one expects Al$^{+}$ to be dominant and
Al$^{2+}$   to be essentially absent in H$^{\circ}$ regions. The observations  
indicate that Al$^{2+}$ is present in DLAs   at the same radial  velocity   
of low   ions,
but at different radial velocity of high ions such as C$^{+3}$ and Si$^{+3}$
   (Lu et al. 1996, Prochaska \& Wolfe 1999, Wolfe \& Prochaska 2000).
Clearly, 
Al$^{2+}$ is a tracer of moderately ionized gas
associated to the neutral phase and the study of the
 Al$^{2+}$/Al$^{+}$ ratio must bear important information on  
the effects of ionization   on the derived abundances.

In a previous work we have performed
detailed ionization   calculations  relevant to  
 the ionized layer directly
exposed to the ionizing radiation field 
(Howk \& Sembach 1999; hereafter Paper II). 
In that work
the photoionization calculations  
have been stopped at the point in which the local ionization fraction
of neutral hydrogen climbs above 10\%.
In the present work we first study the Al$^{2+}$/Al$^{+}$ ratio 
from the observational point of view and  we then perform photoionization
calculations aimed at modeling this ratio.
Rather than analyzing individual absorbers, we study the general behavior of the
Al$^{2+}$/Al$^{+}$ ratio in a relatively large sample of DLAs.
In our   computations 
we  consider the possibility that
the line of sight can intersect both neutral gas and partially ionized gas.
At variance with Paper II,  
 we also perform photoionization calculations 
stopped at fixed   $N$(H$^{\circ}$) values.

\section{ Analysis of observational data. }
In  Table 1 we present a compilation of   aluminium column densities 
in DLAs collected from the literature.  
References to the original works are reported in column 6. 
Only measurements performed on high resolution spectra
obtained with 10-m class telescopes have been included
in the list.   
Most of   the data have been obtained with the Keck HIRES
spectrograph and analyzed by only a few authors. The collected
data set is therefore quite homogeneous from the observational point of view.
Nevertheless, we paid special attention  on the reliability of the published
column densities. In particular,  we revised the
available data for rejecting cases suspected to be affected by saturation or by
telluric contamination.  In addition, we
investigated the velocity distribution of the absorption profiles. 
In fact the
comparison of the column densities of Al$^{+}$,
Al$^{2+}$ and other low ions that we perform here makes sense only if these ions
have similar velocity profiles, suggesting that they are physically related.  
 
The Al$^{+}$ column densities
 are   derived from the transition at $\lambda_{\rm lab}167.0$\,nm
which is often saturated. In fact, only  lower limits  are reported 
in the literature for a large fraction of cases. 
We have, in addition, found three more cases of suspected saturation from
a revision of the absorption profiles.  
We treat also these three cases as lower limits even if they are quoted as
real measurements in the original works (see note $d$ in the table). 

The Al$^{2+}$  column densities are derived from the transitions at
$\lambda_{\rm lab}185.4$\,nm and
$\lambda_{\rm lab}186.2$\,nm
which are not affected by saturation effects. 
Our re-analysis of the velocity profiles confirms that 
Al$^{2+}$ has, in general,  the same velocity
distribution of the low ions (Wolfe \& Prochaska 2000).  However, we found some exceptions
which are marked with
the  explanatory notes $f$ and $g$ in Table 1.   
Typically, we found evidence for an  additional Al$^{2+}$ absorption component 
displaced from the zero velocity component seen in Al$^{+}$
and/or other  low ions.  
In most of these cases the equivalent width of this extra   absorption
is negligible compared with the total Al$^{2+}$ absorption. 
The bulk of   Al$^{2+}$
 originates in the same velocity range as  Al$^{+}$ and the other
low ions.

For the absorber at $z=3.3901$
towards QSO 0000-263 we found evidence for    telluric contamination of
the Al$^{2+}$ 185.4 nm profile  from the analysis of a newly obtained spectrum
of this target by means of UVES at VLT (Molaro et al. 2000). 
 In table 1 we report the column
density obtained from the analysis of the 186.2\,nm line, which is unaffected
by contamination.  
Careful inspection of the full set of profiles in the
other DLAs of Table 1 does not reveal evidence for telluric contamination
since  the profiles of the two lines of the doublet are always similar  
when they are both available (almost in all cases). 

\subsection{ Column-density correlation analysis }

\begin{figure}
\plotone{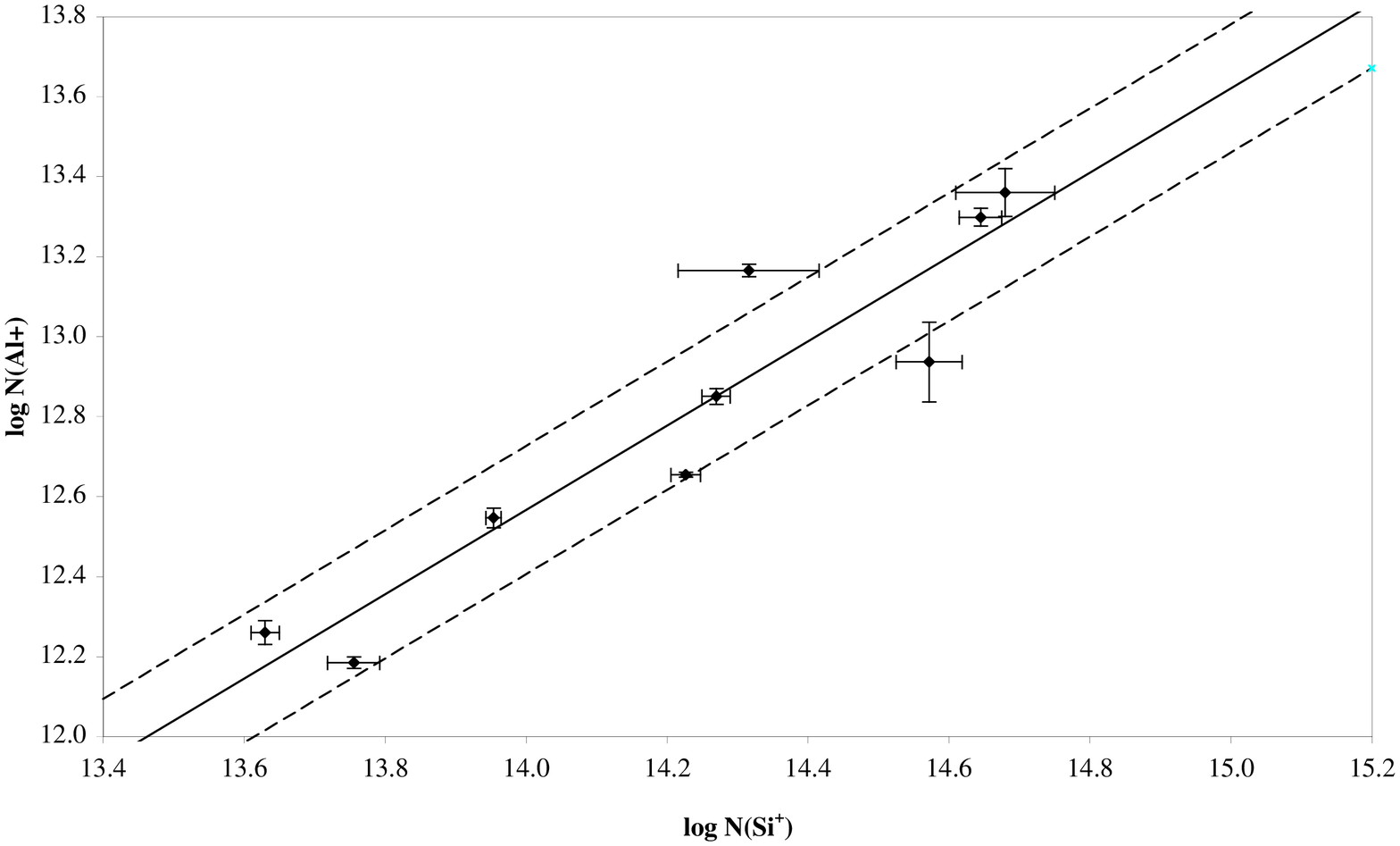}
\caption{Comparison of Al$^{+}$ and Si$^{+}$ column densities
in DLAs. Continuous line: linear regression of the data points.
Dashed lines: $\pm 1 \sigma$ dispersion of the linear regression.
}
\label{Al2&Si}
\end{figure}

While Al$^{+}$ and other singly ionized species with IP $< 13.6$ eV 
can be present in the inner parts of H$^{\circ}$ regions,  the production of
Al$^{2+}$ requires photons with $h\nu > 18.8$ eV
which cannot as easily penetrate large H$^{\circ}$ column densities. 
A careful analysis 
should therefore reveal  a different behavior between  Al$^{2+}$ and
singly ionized species. Since the radial velocity analysis does not show evidence for a distinct behavior,  
we investigated the  column density behavior of such species. 
For two species that are present in roughly constant proportions
in the same region we expect to find a linear
correlation with  slope  of unity between their logarithmic column densities.  
The  identification of  such a correlation would support a common origin of 
the different species. The lack
of such a correlation could be ascribed   to  
differences in the   abundance/depletions patterns
or to differences in the ionization properties. 
By analyzing ions of the same elements,   such as 
Al$^{2+}$ and Al$^{+}$, there can be no differences in the intrinsic abundances.
Studying such ions, therefore, gives information on the ionization state
of the gas, and a slope differing from unity would indicate that ionization properties
do  affect the observed column densities.  
Unfortunately, only a modest number of simultaneous determinations of 
Al$^{+}$ and Al$^{2+}$ in   DLAs  are currently known
(Table 1).  
To by-pass this limitation we compared the column densities
of both Al$^{+}$ and Al$^{2+}$ ions with those of  Fe$^{+}$ and Si$^{+}$, for which a large
number of measurements are  instead available.  In practice,  we performed a linear regression 
analysis of the column densities of each
possible combination of pairs of the species 
Al$^{+}$, Al$^{2+}$, Si$^{+}$, and Fe$^{+}$.
The results are summarized in Table 2, where we give for each pair  the
number of data points available,  $n$, the correlation coefficient, $r$,
the slope, $m$, and the intercept, $q$, resulting from the linear regression. 
In Fig. \ref{Al2&Si} we show the  results for Al$^{+}$ and Si$^{+}$.

One can see from Table 2  
that singly ionized species, including Al$^{+}$ are well correlated with each other and
have   slope  of unity within the 
errors\footnote{ 
Chemical evolution and dust effects  are hidden within the dispersion
of the correlations. At first sight it is surprising that such effects
do not introduce   a large scatter. 
However it is possible to show that changes in metallicity level
and dust content tend to compensate each other
since relative abundances scale with the
dust-to-metals ratio   (Eq. 17 in Paper I) and at the same
time 
metallicity   and dust content are   correlated  in DLAs (Fig. 2 in Paper I). 
On the other hand, the intrinsic spread of relative abundances due to
 chemical evolution  may be modest owing to
relatively limited range of look-back time explored in DLAs
(most absorbers are found at $z \simeq 2.5$). 
}. 
 This fact,
together with the similarity of the velocity profiles, suggests that all these species
originate in the same region. 
On the other hand, pairs of species including
Al$^{2+}$ show a modest correlation coefficient and a slope significantly lower
than 1. In addition, such pairs have larger dispersions than the
corresponding pairs with Al$^{+}$. For instance $\sigma$(Al$^{2+}$,Si$^{+}$)
$\simeq$ 2 $\sigma$(Al$^{+}$,Si$^{+}$). 
This distinct behavior of Al$^{2+}$
cannot be attributed to   differences in the abundances or dust depletions
properties of Al, Fe and Si. If such effects were important, 
they would also tend to cancel the correlations between  
Al$^{+}$ and Fe$^{+}$ and between Al$^{+}$ and Si$^{+}$,
which are instead clearly detected. 
Since the distinct behavior of Al$^{2+}$ is not due to intrinsic variations
of the abundances/depletions, it must be due to the ionization properties
of DLA clouds.  This observational result can be interpreted in two ways:

\begin{enumerate}

\item
 The Al$^{2+}$ originates in the same H$^{\circ}$ region as the bulk of the
low-ionization lines.  The local value of the Al$^{2+}$/Al$^{+}$
  ratio varies
among DLA clouds.

\item
The Al$^{2+}$  originates in a region distinct from, but physically
related to, the H$^{\circ}$  region where the bulk of the singly-ionized species
reside.  The relative column density contribution from the two regions
varies among DLA clouds. 

\end{enumerate}

Studies of the ionization properties of DLAs usually consider only
the first of the two possibilities. Here we will also consider the second one. 
A separate Al$^{2+}$ region may in fact exist if 
the photons with $h\nu > 18.8$ eV required for producing Al$^{2+}$ 
are not available in the self-shielded parts of   the  H$^{\circ}$  region.
Since the velocity structure of  Al$^{2+}$ and of low ions   are generally similar,
the Al$^{2+}$ and H$^{\circ}$ regions must be physically connected. 
A cloud structure consisting of a partially ionized   Al$^{2+}$-bearing interface bordering
a H$^{\circ}$  region  opaque to $h\nu > 13.6$ eV photons    
satisfies the above requirements.

\subsection{The  Al$^{2+}$/Al$^{+}$ ratio in DLAs}

The   ratio Al$^{2+}$/Al$^{+}$   can be used to constrain the ionization conditions
and/or ionization structure of DLAs.   From the observational point of view
we are   able to measure   the
column density ratio $\cal R$(Al$^{2+}$/Al$^{+}$) $\equiv$ $N$(Al$^{2+}$)/$N$(Al$^{+}$), which gives
information integrated along the line of sight. 
Direct measurements of  $\cal R$(Al$^{2+}$/Al$^{+}$)
 are currently possible only in a few cases,
mainly because reliable Al$^{+}$ measurements are quite rare (Table 1). 
For this reason we take advantage of the existence of the 
$\log N({\rm Al}^{+})$ -- $\log N({\rm Fe}^{+})$ and
$\log N({\rm Al}^{+})$ -- $\log N({\rm Si}^{+})$ correlations 
to indirectly estimate  $N$(Al$^{+}$)  from Fe$^{+}$ and Si$^{+}$ measurements.
The indirect determinations based on
Si$^{+}$ measurements are displayed 
in  Table 1, column 7.
References to the original Si$^{+}$ measurements are the same given
in column 6.
These indirect estimates of $N$(Al$^{+}$)   yield   results consistent
with all the    available Al$^{+}$ data
(i.e., 9 determinations and also 9 lower limits). 
Similar results are obtained when Fe$^{+}$ is used to track Al$^{+}$
(see more details in Section 2.4). 
Because the correlation with    
$N$(Si$^{+}$) has  lower
dispersion, lower slope error, and lower intercept error  than the correlation 
with  $N$(Fe$^{+}$) (Table 2) we use 
the Si$^{+}$-based results given in Table 1
in the rest of the paper, unless otherwise specified.

In the last column of Table 1 we list the values of the aluminium ionization ratio 
in DLAs obtained by using indirect and, when possible, direct measurements of Al$^{+}$. 
The ionization ratio shows a spread of more than one order of magnitude,
with values ranging
from   $\cal R$(Al$^{2+}$/Al$^{+}$) $\simeq$ 0.03  up to $\simeq$ 0.6, and a
 median value $\simeq$ 0.2.  
Values of  $\cal R$(Al$^{2+}$/Al$^{+}$) as high as 0.6 are detected also in
the sub-sample for which $N$(Al$^{+}$) is measured directly. 
The presence of high fractions of doubly ionized aluminium   suggests that
Al$^{+}$ is not the only ionization state of aluminium present in significant amounts
in DLAs. However, 
it is unsafe to use    
 $\cal R$(Al$^{2+}$/Al$^{+}$), which is a quantity integrated along the line of sight,
as an indicator of the local ionization ratio. In fact,   
  singly and doubly ionized aluminium may
arise in distinct regions, in which case   $\cal R$(Al$^{2+}$/Al$^{+}$)
 would reflect  the relative contribution of such regions to the  column densities, 
rather than the local ionization ratio.

The $\cal R$(Al$^{2+}$/Al$^{+}$) ratio does not show any trend with absorber redshift.
In the   interval $2.0 \leq z_{\rm abs} \leq 2.5$, where most of the
measurements are concentrated,  the ratio shows the full spread of a factor of 20.
Since metagalactic effects  are expected to have a smooth variation with redshift,
the large spread in a relatively narrow redshift interval suggests that the ionization fraction is 
severely influenced by local effects.   
The metagalactic radiation field may play  a role in determining the
ionization balance, but must be modulated by some local effect.  
Such modulation may result from   emission/absorption  of radiation  
internal to the DLAs or from large changes in cloud density.

\subsection{ Abundances versus   Al$^{2+}$/Al$^{+}$  ionization ratio }
The presence of high fractions of Al$^{2+}$   with same velocity distribution of 
low-ionization  might
question the reliability of abundance determinations in DLAs.
If  abundance  measurements are strongly affected by ionization
conditions, we would expect that they show some dependence on $\cal R$(Al$^{2+}$/Al$^{+}$).  
For instance, the models calculations performed in Paper II
predict that    [Zn/H] and [Si/H] measurements\footnote{
We adopt the usual convention 
[X/Y] $= \log [ { N({\rm X}) \over N({\rm Y}) } ] - \log ( {\rm X} / {\rm Y} ) _{\sun}$. 
}
 can be easily
overestimated by one order of magnitude 
when  $\cal R$(Al$^{2+}$/Al$^{+}$)   $\geq$ 1/10 for the relatively low
ionization parameters discussed in that work.  
The collection of $\cal R$(Al$^{2+}$/Al$^{+}$)   values presented in Table 1
allows us to probe  if such effects are indeed present, at least for 
elements commonly measured in DLAs. 
We   performed a linear regression analysis of [Si/H] and [Si/Fe] against
$\log \cal R$(Al$^{2+}$/Al$^{+}$). 
The [Si/H] analysis yields 
correlation coefficient $r=0.29$, dispersion $\sigma=0.46$, 
and slope $m=0.37 \pm 0.30$  (18 data points). 
The predicted increase of [Si/H] at high $\cal R$(Al$^{2+}$/Al$^{+}$), if present, 
is not significant from the statistical point of view. 
Similar results are obtained from the study of [Si/Fe],
a ratio which is    expected to be moderately sensitive to ionization effects
(Paper II). 
The    [Si/Fe] analysis   yields 
$r=0.05$,  $\sigma = 0.20$, and $m=-0.03 \pm 0.13$ (19 data points). 
In this case the null result is even more clear.
The lower dispersion than in the case of the Si/H ratio 
is probably due to the fact that  
variations of the metallicity and of the dust-to-gas ratio tend to cancel when
we  consider a relative abundance such as Si/Fe. 
These results suggest that ionization corrections  are not severe, or at least that they
do not show a strong, obvious dependence on $\cal R$(Al$^{2+}$/Al$^{+}$).

\subsection{Al$^{2+}$/Al$^{+}$ ratio versus H${^\circ}$ column density}
 
\begin{figure}
\plotone{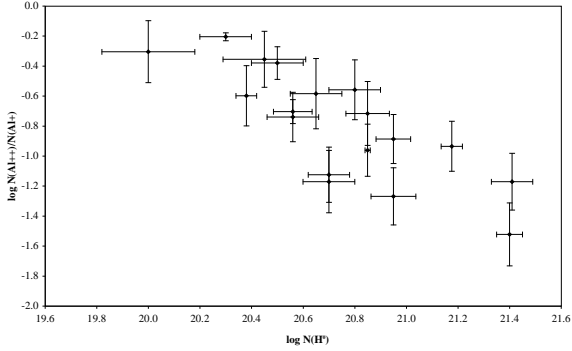}
\caption{Ionization ratio 
 $\cal R$(Al$^{2+}$/Al$^{+}$)
versus neutral
hydrogen column density in DLAs. Data points taken from Table 1.
}
\label{iAl&HI}
\end{figure}

In Fig. \ref{iAl&HI} we plot $\cal R$(Al$^{2+}$/Al$^{+}$)  versus the absorber
H$^{\circ}$ column density. 
The data points  
show a general decrease of   $\log \cal R$(Al$^{2+}$/Al$^{+}$) 
with increasing log\,$N$(H$^{\circ}$).  
A linear regression analysis yields an anti-correlation with
slope $m = -0.81  \pm 0.15$ and intercept $q = 16.0 \pm 3.1$.
Even if the   correlation coefficient is
not very high (Pearson's $r=-0.80$), 
the probability of the null hypothesis (of no correlation) is
$5.9 \times 10^{-5}$.
The highest value of the ratio, $\cal R$(Al$^{2+}$/Al$^{+}$)   $\simeq  0.6$,
is found at  $N$(H$^{\circ}$) $\simeq$ 10$^{20.3}$ cm$^{-2}$, while the 
lowest value, $\cal R$(Al$^{2+}$/Al$^{+}$)   $\simeq 0.03$,
is found at  
$N$(H$^{\circ}$) $\simeq$ 10$^{21.4}$ cm$^{-2}$.

The above result   is based in large part on indirect Al$^{+}$ determinations obtained from
 Si$^{+}$ measurements. However, the 
$\log \cal R$(Al$^{2+}$/Al$^{+}$)  -- log\,$N$(H$^{\circ}$)
anti-correlation is confirmed   when 
Fe$^{+}$ is used to infer Al$^{+}$ column densities. In fact, 
the slope and intercept  derived in this case --- 
$m = -0.78  \pm 0.16$   and  $q = 15.2 \pm 3.3$ ---
are very similar
to the above given values, and the probability of the null hypothesis
equals $8.6 \times 10^{-5}$.

The lack of DLAs at the bottom-left
and top-right corners of Fig. \ref{iAl&HI} could be   due, in principle,
to the impossibility of detecting 
Al$^{+}$ and/or Al$^{2+}$ below their observational limits.  
However, we have verified that this is not the case.
In fact,  the strength of the Al$^{+}$ 167.0 nm transition
and of the numerous Si$^{+}$ and Fe$^{+}$ transitions used for the indirect Al$^{+}$
determinations guarantee that Al$^{+}$ would be easily detected down to much
lower column densities than observed.  In the
case of Al$^{2+}$, the detection limit with  Keck+HIRES is $\simeq 10^{12}$ cm$^{-2}$.
In the sample of Table 1, only one case out of 21 measurements has such a low value.
Of the remaining $N$(Al$^{2+}$) data, 9 are  0.5 dex above the detection limit and 11 are 1 dex 
above the limit. We conclude that the anti-correlation is not an artifact induced by  
observational limitations. 

In addition, we do not have reasons to believe that
the observed trend results from a selection bias. The most common bias
considered to affect the population of DLAs is dust obscuration of the
background QSO (Pei et al. 1991). In fact, this effect
may be responsible for the apparent anti-correlation between [Zn/H]
and $N$(H$^{\circ}$) found by Boiss\'e et al. (1998), if lines of sight of
high metallicity (dust content) and high column density indeed obscure the background QSO.
However, the same effect should not be relevant for the anti-correlation
reported here given the fact that
 the ratio $\cal R$(Al$^{2+}$/Al$^{+}$) is independent of
metallicity.

The existence of an empirical anti-correlation 
$\log N$(H$^{\circ}$) -  $\log \cal R$(Al$^{2+}$/Al$^{+}$) implies that
the neutral hydrogen column density can be used as an indirect estimator
of the  ionization state of the gas. In this way one can search for ionization
effects in DLAs  without available Al measurements.
 Given the importance of Zn in DLA studies we searched for a possible
dependence of [Cr/Zn] and [Fe/Zn] on $\log N$(H$^{\circ}$).
A linear regression through the 26   [Cr/Zn] measurements available  yields
a slope $m=-0.03 \pm 0.11$ and a correlation coefficient $r=0.06$. 
Very similar results are obtained from the analysis of the 22 [Fe/Zn]   available 
determinations: $m=-0.04 \pm 0.13$ and $r=0.07$. 
In addition,   there is no trace of an increased dispersion at low $N$(H$^{\circ}$)
for any of the two ratios.   
These empirical results suggest that Zn ionization effects are probably
negligible in DLAs.

If Al$^{2+}$/Al$^{+}$ is relatively constant throughout the cloud,
the $\cal R$(Al$^{2+}$/Al$^{+}$) ratio should be approximately constant with $N$(H$^{\circ}$).
The decrease of $\cal R$(Al$^{2+}$/Al$^{+}$) with  $N$(H$^{\circ}$) 
can be interpreted in two ways, depending on which of the two possible origins of Al$^{2+}$
mentioned at the end of Section 2.1 is more appropriate. 
(1) If Al$^{2+}$ is co-spatial with
the rest of low-ionization species, the anticorrelation 
is consistent with
a reduction of the gas ionization level  with increasing H$^{\circ}$
self-shielding. 
(2) If Al$^{2+}$ originates in a partially ionized interface 
bordering a neutral region, 
then we expect 
the column density of the ionized interface and of the neutral region to be unrelated 
and therefore
$N$(Al$^{2+}$) to be  independent of $N$(H$^{\circ}$).  
On the other hand,  we expect Al$^{+}$ and other species of low ionization
to scale with H$^{\circ}$ owing to a common origin in the neutral region. 
The decrease of $\cal R$(Al$^{2+}$/Al$^{+}$) with $N$(H$^{\circ}$) is
naturally explained from the fact that $N$(Al$^{2+}$) does not scale with $N$(H$^{\circ}$)
while $N$(Al$^{+}$)  does.

\section{ Model calculations }

On the basis of the observational results discussed above, we consider two
possible origins for   species of low and intermediate ionization:
(1) regions completely opaque to   photons with $h\nu > 13.6$\,eV 
and/or
(2) regions   partially transparent to ionizing photons. 
Neutral and singly ionized species can arise in both regions, but
Al$^{2+}$ can only be present in   type-2 regions. 
The general idea is that type-2 regions are  
 the  photoionized  envelopes of type-1 regions.
The similarity of the velocity profiles of Al$^{2+}$ and singly ionized species
is naturally explained in this way.
On the other hand,
radial-velocity studies show a general misalignment of C$^{+3}$ and Si$^{+3}$ profiles
relative to Al$^{2+}$ profiles, suggesting that the bulk of high-ionization species 
originates elsewhere  (Wolfe \& Prochaska 2000). 
We expect therefore that type-2 regions are mildly ionized and have
low    Si$^{+3}$/Si$^{+}$ fraction. 

In the Appendix we derive the  expressions that allow us to compute
ionization ratios and abundance ionization corrections in the framework
of the two-region model. 
These expressions do not depend on the metallicity, provided type-1 and type-2
regions have equal metallicity in any given DLA system.
The relative contribution of the two regions along the line of sight is
specified by the parameter $N_1/N_2$, which is the ratio of the 
total column densities in region 1 and 2, respectively. 
The possibility that Al$^{2+}$ and  the species of low ionization 
 originate together in a single layer   is   included in our treatment 
as the special case  $N_1/N_2=0$. In fact, the single layer should be
a partially ionized region of type 2 given the presence of significant
fractions of Al$^{2+}$.

In order to model the ionization properties of the gas
we assume that the DLA regions under study are embedded in a ionizing radiation field
of given spectrum and intensity.  
In type-1 regions we only consider the contribution of species which are
dominant ionization states in HI regions. In type-2 regions we compute the
ionization fractions by means of photoionization equilibrium calculations.
For this purpose we used   the CLOUDY code (v90.04; Ferland 1996, Ferland et al. 1998)
assuming plane-parallel geometry with ionizing radiation incident on one side.
We consider 
two possible types of   radiation fields: 
a hard, QSO-dominated spectrum representative of the radiation field 
external to  the DLAs at $z=2$
(Haardt \& Madau 1996; Madau, Haardt  \& Rees 1999)
and a soft,  stellar-type spectrum ($T_{\rm eff}$=33,000 K; Kurucz 1991)
representative  of  the  internal radiation field
or of an external field dominated by starlight from galaxies
(e.g. Steidel, Pettini \& Adelberger 2000). 
In both cases the intensity of the   field is specified by the  ionization parameter
$U \equiv \Phi({\rm H}) / c \, n_\mathrm{H}$ , where $\Phi$(H)
is the total surface flux of ionizing photons (cm$^{-2}$ s$^{-1}$)
and $n_\mathrm{H}$ the hydrogen particle density (cm$^{-3}$). 
We refer to  Paper II for  more details on the model  assumptions and  on the
adopted radiation fields.  
In the  rest of this section we present model predictions of the 
ionization ratios ${\cal R}$(Al$^{2+}$/Al$^{+}$) and of 
the abundance ionization corrections.

\subsection{ Aluminium ionization ratio  }

\subsubsection{ Soft-continuum, two-regions model (S2 model) }

The photoionization calculations indicate that with the soft,
stellar-type spectrum  the 
  H$^{\circ}$ column density  of the ionized layer, $N_2$(H$^{\circ}$), cannot attain
the high values
typical  of DLAs  for a wide range of $U$. In fact,   we find $N_2$(H$^{\circ}$) $\leq 3.2 \times
10^{18}$ cm$^{-2}$ for
$-5.4 \leq \log U\leq -1.0$. 
Therefore, we assume that a neutral region of type 1 is present along the line of sight,
in addition to the Al$^{2+}$-bearing, type-2 region. 
We refer to
this soft-continuum, two-regions model as S2. 

Because the ionized region does not yield a significant contribution to the neutral
hydrogen column density and because  $x({\rm H}^{\circ})=1$
in the neutral region, we take  
$N_1$(H) = $N_1({\rm H}^{\circ})$= $N({\rm H}^{\circ})$ 
as the total column density of the type-1 region,
where  $N({\rm H}^{\circ})$ covers the interval  
$10^{20.2}$ cm$^{-2}$ $ < N({\rm H}^{\circ}) < 10^{22}$  cm$^{-2}$,
representative of DLAs. 
For each   given $N_1$(H)
we let  the  ionization parameter $U$ vary and we use
CLOUDY to determine
the  column density of the type-2 region,
$N_2$(H), and the mean ionization fractions of the species of interest.
We then   estimate 
${\cal R}$(Al$^{2+}$/Al$^{+}$)  by means of   Eq. \ref{AluminiumRatio2}, where we take
$N_1/N_2 = N_1$(H)/$N_2$(H)
and we insert the Al$^{+}$ and Al$^{2+}$  ionization fractions.
In this way we derive model curves of
${\cal R}$(Al$^{2+}$/Al$^{+}$)  versus
$N$(H$^{\circ}$) that can be compared with the observed data points.

\begin{figure}
\plotone{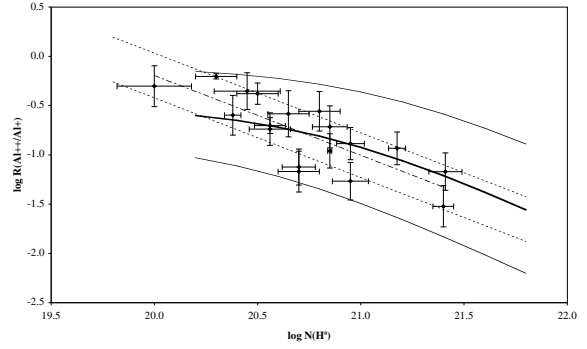}
\caption{ Model predictions
of the $\cal R$(Al$^{2+}$/Al$^{+}$)  ratio versus $N$(H$^{\circ}$)  in DLAs
in the case of a soft, stellar ionizing source (model S2 discussed in Section 3.1.1).  
Solid curves:  $\log \cal R$(Al$^{2+}$/Al$^{+}$) predicted  
at $\log U=-2.6$ (bottom curve), $-2.2$ (thick line), and $-1.7$  (top curve).  
Diamonds: observational data points as in Fig. \ref{iAl&HI};
dashed-dotted line: linear regression through the observed data;
dotted lines: $\pm 1 \sigma$ dispersion of the linear regression.
}
\label{iAlHImodA}
\end{figure}

In   Fig. \ref{iAlHImodA} we show the
results of these calculations for  constant values
of the ionization parameter.
The solutions calculated at $U = 10^{-2.2}$ (thick curve)
are consistent, within the
statistical errors, with the linear regression of the observational data points
(dashed-dotted line).
The corresponding total hydrogen column density of the type-2 layer predicted by the model is
$N_2$(H) $= 8 \times 10^{20}$ cm$^{-2}$. 
The solutions calculated at 
$U_{\rm min}=10^{-2.6}$ and $U_{\rm max}=10^{-1.7}$ (lower and upper thin curves,
respectively)
bracket  the  observational data points.   
For these lower and upper envelope curves we find
$N_2$(H) $= 3.3 \times 10^{20}$ and $2.3 \times 10^{21}$ cm$^{-2}$, respectively.  
The total column densities of type-1 and type-2
regions are roughly comparable ($N_1/N_2 \approx 1$).  	
It is remarkable that the  $\log N$(H$^{\circ}$) - $\log {\cal R}$(Al$^{2+}$/Al$^{+}$)  
anticorrelation can be easily modeled 
with very simple  assumptions and  with a single interval of $U$ values
in the framework of the S2 model.

\subsubsection{ Hard-continuum, single-region model (H1 model) }

\begin{figure}
\plotone{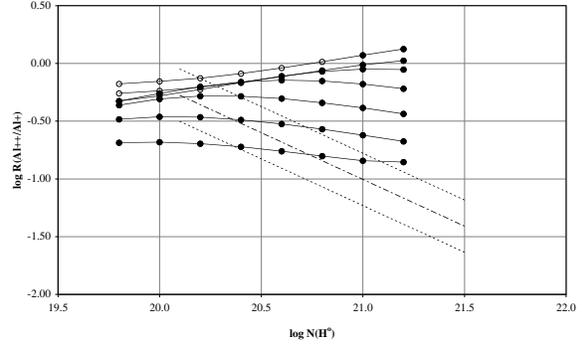}
\caption{  Model predictions
of the $\cal R$(Al$^{2+}$/Al$^{+}$)  ratio versus $N$(H$^{\circ}$)  in DLAs
in the case of a hard, QSO-dominated ionizing source (model H1 discussed in Section 3.1.2).
The continuous curves    have been obtained at constant value 
of the photoionization parameter, $U$. Curves from bottom to top
correspond to $\log U = -4.8, -4.2,   -3.6, -3.0,  -2.4, -1.8$
and $-1.2$, respectively. Filled and empty symbols
represent solutions for which $\cal R$(Si$^{+3}$/Si$^{+}$) $\leq -0.5$ dex
and   $\cal R$(Si$^{+3}$/Si$^{+}$) $> -0.5$ dex, respectively.  
Dashed-dotted line: linear regression through the observed data points.
Dotted lines: $\pm 1 \sigma$ dispersion of the linear regression
trough the observed data.  
}
\label{iAlHIQSOa}
\end{figure}

With the QSO-dominated field the H$^{\circ}$  column density of the ionized layer
can attain the high values typical of DLAs if 
$U$ is sufficiently high. As a consequence, the presence of a 
of type-1 region is not required 
in order to explain the observed $N$(H$^{\circ}$) values.  
In this case we assume that all the low and intermediate ionization species
originate in a single, partially ionized layer.
We refer to
this hard-continuum, one-region model as H1. 
We stopped the calculations at a series of $N$(H$^{\circ}$) 
values representative of DLAs, rather than at a given threshold of ionization fraction.  
Our treatment of this case is therefore quite similar to those of
Lu et al. (1995) and Prochaska \& Wolfe (1996). 
At variance with the S2 model, in the H1 model there is residual ionized gas,
and hence Al$^{2+}$, 
throughout the interior of the cloud  
due to the strong high-energy tail of the extragalactic Haardt \& Madau spectrum. 
In Fig. \ref{iAlHIQSOa} we show  the resulting  ${\cal R}$(Al$^{2+}$/Al$^{+}$)
ratios plotted versus $N$(H$^{\circ}$) at constant values of $U$.
We   estimated  ${\cal R}$(Al$^{2+}$/Al$^{+}$) from Eq. \ref{AluminiumRatio2}
by taking $N_1/N_2=0$.  
When $U$ is too low there are difficulties with the thermal solution in the
CLOUDY computation since the temperature falls well below 1000 K before the
total $N$(H$^{\circ}$) is reached. This explains the lack of solutions 
below $\log U = -4.8$ in Fig. \ref{iAlHIQSOa}.  
On the other hand, some of the   solutions   tend to overproduce    
Si$^{+3}$ when  $U$ is too high. As an example,  in Fig. \ref{iAlHIQSOa} we indicate with
empty circles the solutions for which   
${\cal R}$(Si$^{3+}$/Si$^{+}$) $> -0.5$ dex. 
The intrinsic ${\cal R}$(Si$^{3+}$/Si$^{+}$) is probably much lower than
this conservative limit given the very different velocity structure of 
 Si$^{3+}$ and Si$^{+}$. 

With the H1 model the curves predicted  at constant $U$
 have very different    -- in some cases opposite -- slope from that
of the anticorrelation $\log N$(H$^{\circ}$) - $\log {\cal R}$(Al$^{2+}$/Al$^{+}$)
(Fig. \ref{iAlHIQSOa}). 
When all the $U$ values are considered,
the solutions tend to fill the
plane $\log N$(H$^{\circ}$) - $\log {\cal R}$(Al$^{2+}$/Al$^{+}$)  
without any preference for the regions populated by observational data.  
To reproduce the anti-correlation it is necessary to impose  very specific
constraints to the input parameters. 
For a given  $N$(H$^{\circ}$), there is an allowed interval
of $U$ values such that the predicted $\log {\cal R}$(Al$^{2+}$/Al$^{+}$) ratios
overlap the observed ones. 
For instance, at $\log N$(H$^{\circ}$) = 20.4 we must require that
$-4.8 \lsim \log U \lsim -3.6$ in order to obtain   ratios
that lie within $\pm 1 \sigma$ of the regression to the observed data points
(Fig. \ref{iAlHIQSOa}). In addition,  the ionization parameter must
decrease, on the average, while $N$(H$^{\circ}$) increases. 
For instance, we must require that $U \propto N({\rm H}^{\circ}) ^{-1.5}$
in order to reproduce the slope of the observed anticorrelation
in the range $20.4 \leq \log {\rm N}({\rm H}^{\circ})\leq 20.8$. 
This is at variance with the results of the S2 model, for which 
all the observed data points are easily matched by adopting a single
interval of $U$ values.

The use of one-side illuminated clouds is inherent to the design of CLOUDY, but may be
slightly inappropriate for the external, hard ionizing spectrum case.
According to Prochaska \& Wolfe (1996) 
1-side  calculations tend to give a lower   degree of ionization 
than 2-side calculations. 
We have estimated this effect by doubling the column densities of the
1-side calculations. 
With this type of estimate we do not find
significant changes in the results of the present work, including the abundance corrections
discussed below.

\subsection{ Abundance ionization corrections  }

\begin{figure}
\plotone{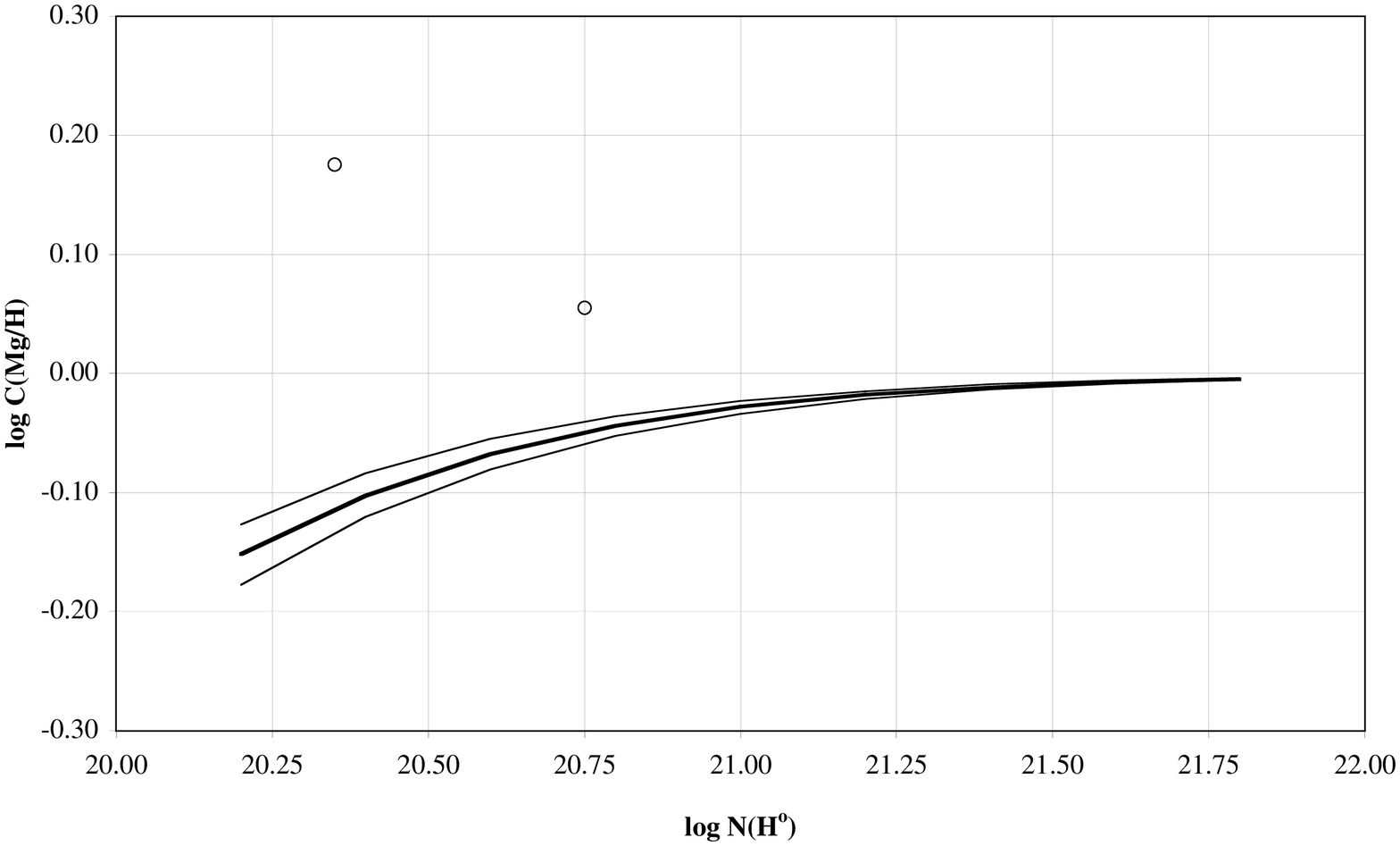}
\caption{
Ionization correction terms for  [Mg/H] measurements  in DLAs.   
Continuous curves:   predictions for the S2 model discussed in Section 3.1.1
calculated at constant values of the ionization parameter $U$.
Thick curve: $\log U = -2.2$. Thin curves:  $\log U = -1.7$ (bottom)
and $\log U = -2.2$ (top).  
Empty circles: predictions for the H1 model discussed in Section 3.1.2
calculated at $\log U =-4.2$ (left symbol) and   $\log U =-4.8$
(right symbol). 
} 
\label{cMgH}
\end{figure}

\begin{figure}
\plotone{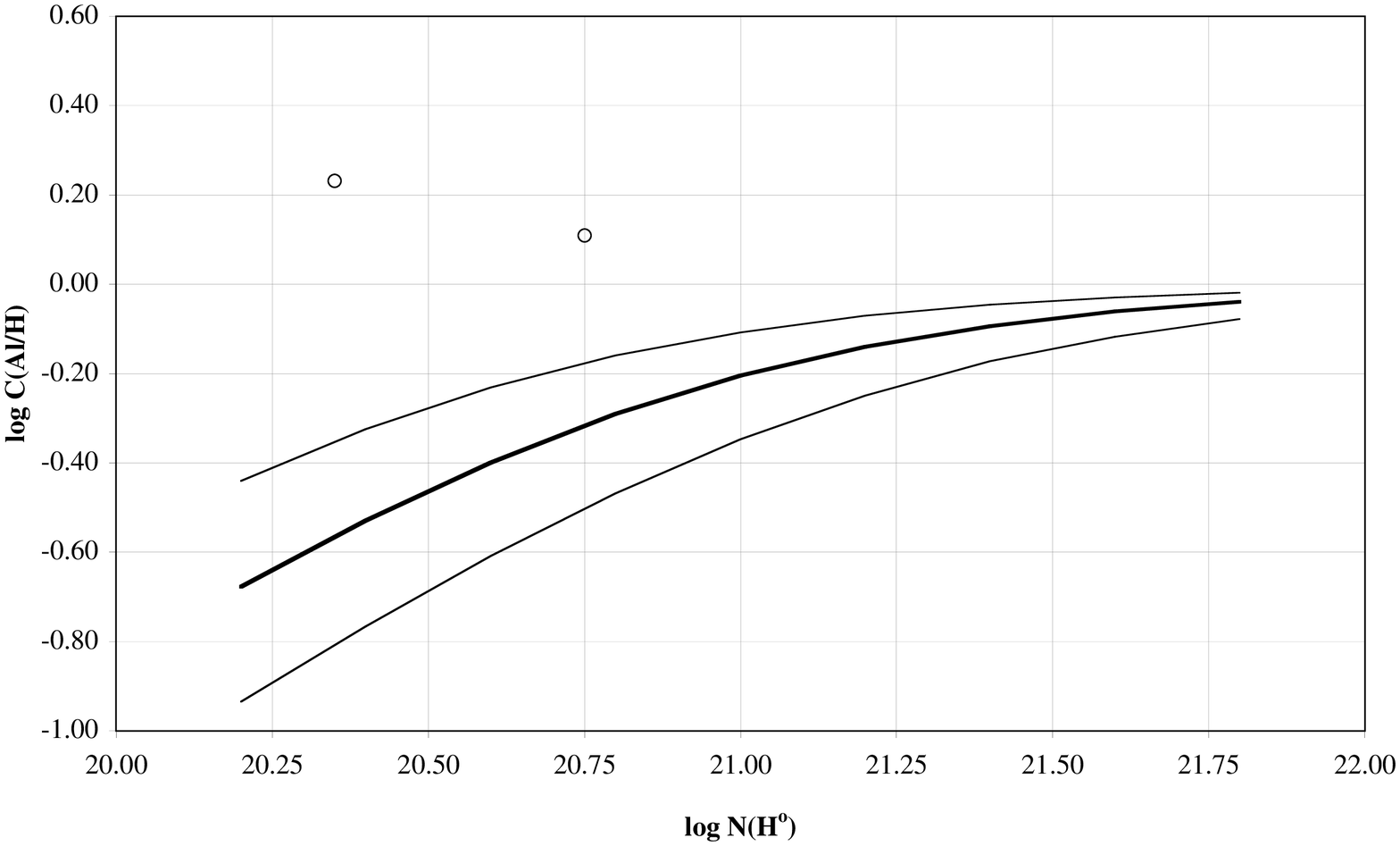}
\caption{
Ionization correction terms for  [Al/H] measurements.   
Legend as in Fig. \ref{cMgH}.} 
\label{cAlH}
\end{figure}

\begin{figure}
\plotone{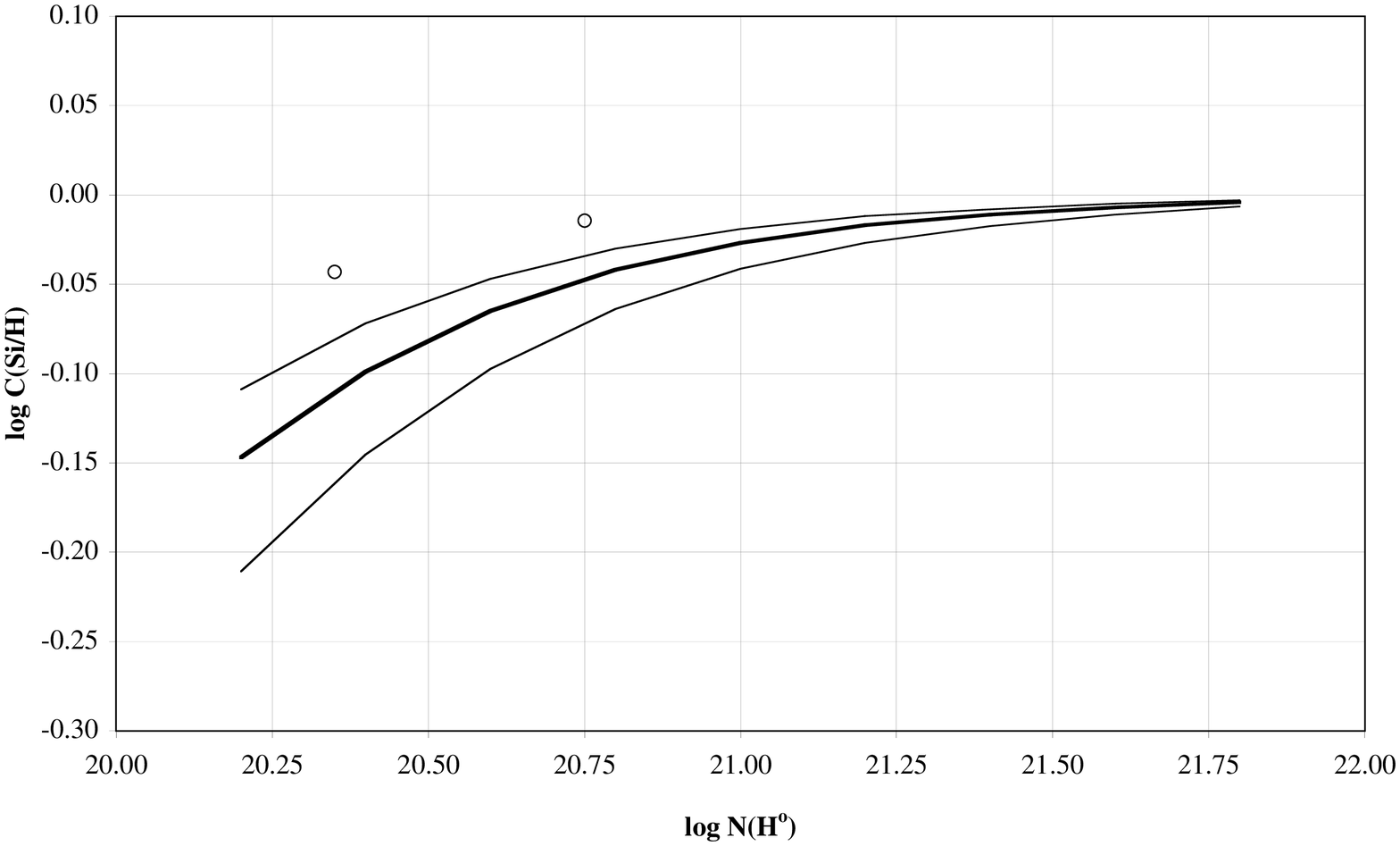}
\caption{ 
Ionization correction terms for  [Si/H]  measurements.
Legend as in Fig. \ref{cMgH}.
 } 
\label{cSiH}
\end{figure}

\begin{figure}
\plotone{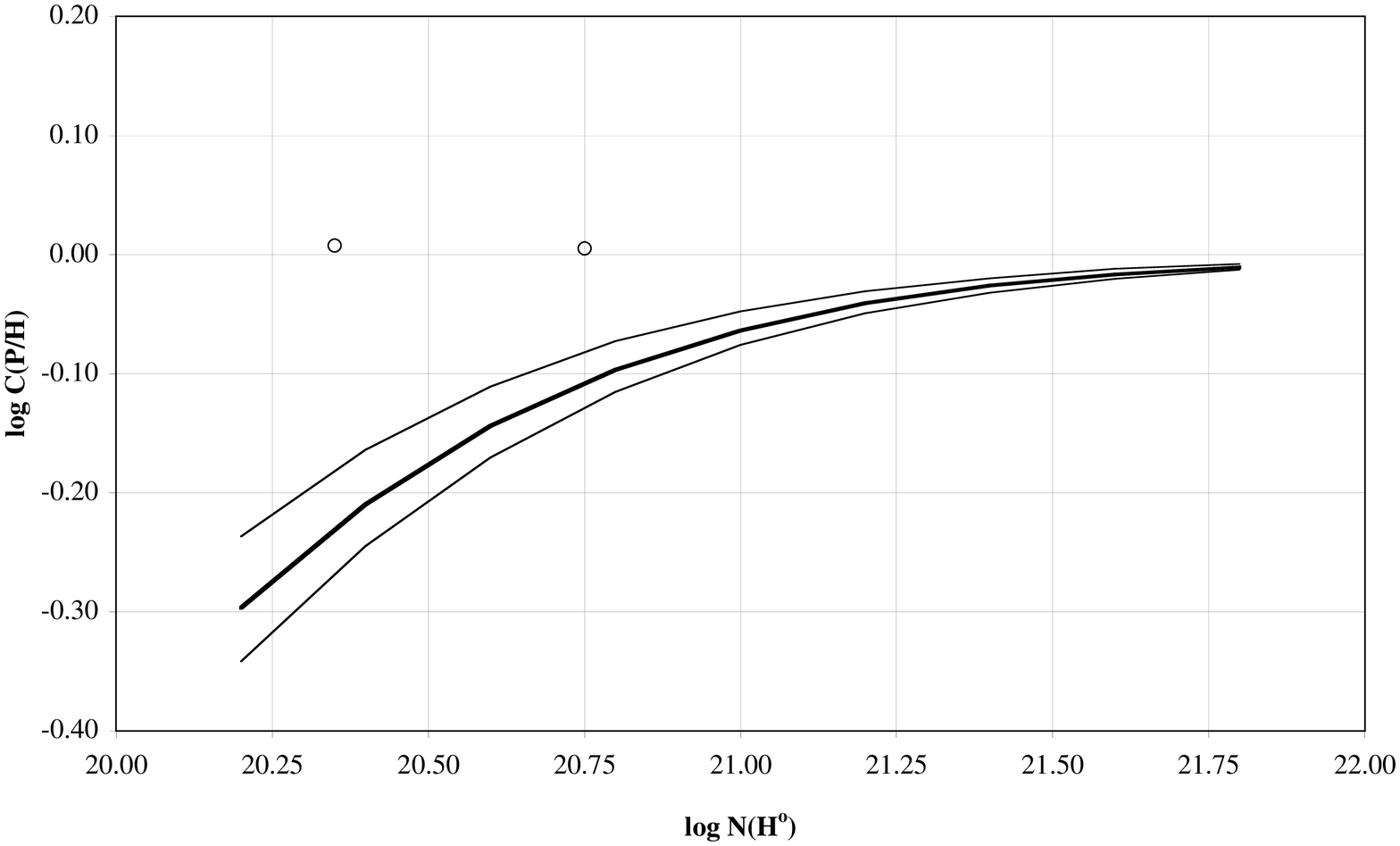}
\caption{ 
Ionization correction terms for  [P/H]  measurements.
Legend as in Fig. \ref{cMgH}.
 } 
\label{cPH}
\end{figure}

\begin{figure}
\plotone{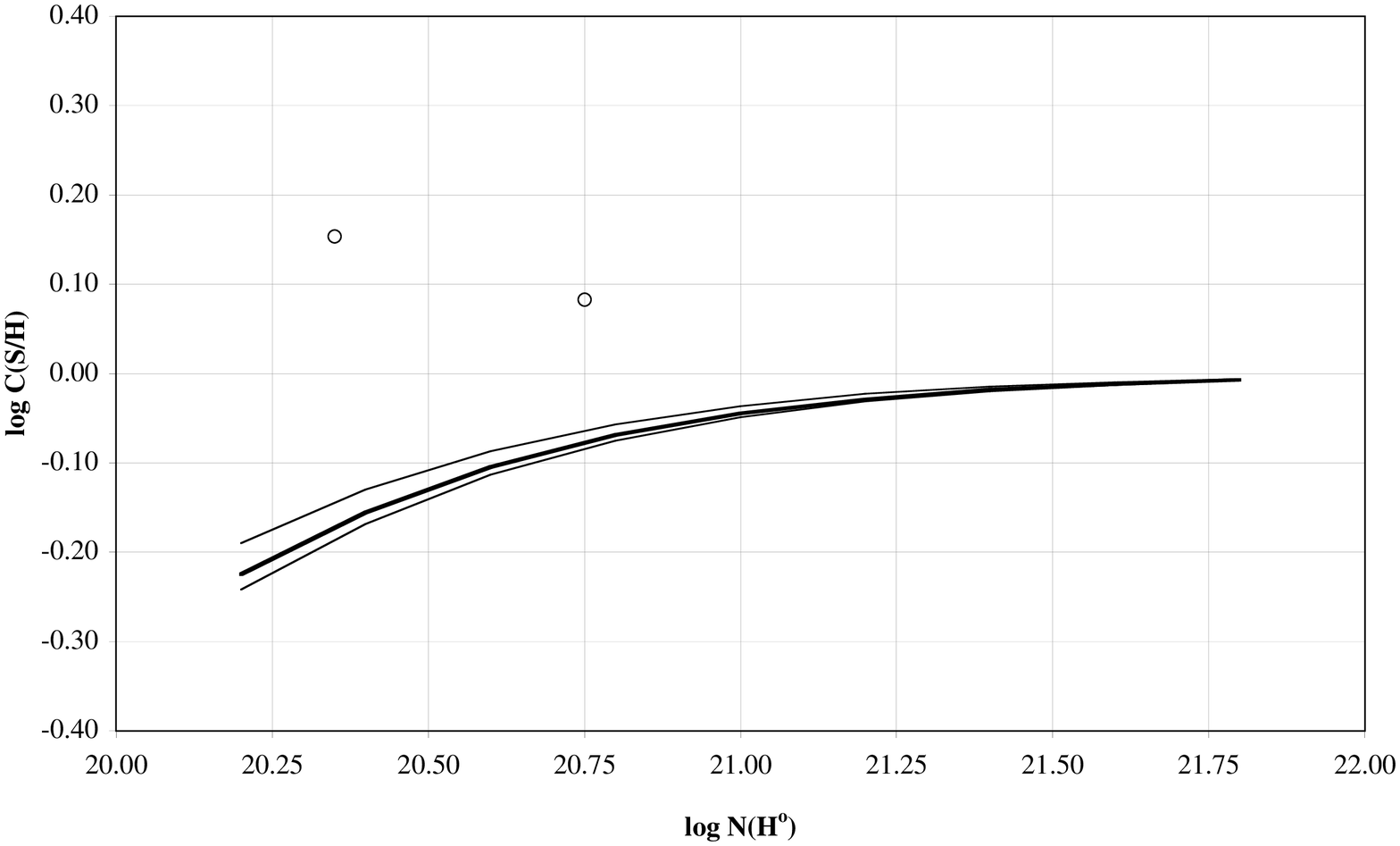}
\caption{
Ionization correction terms for  [S/H] measurements.
Legend as in Fig. \ref{cMgH}.} 
\label{cSH}
\end{figure}

\begin{figure}
\plotone{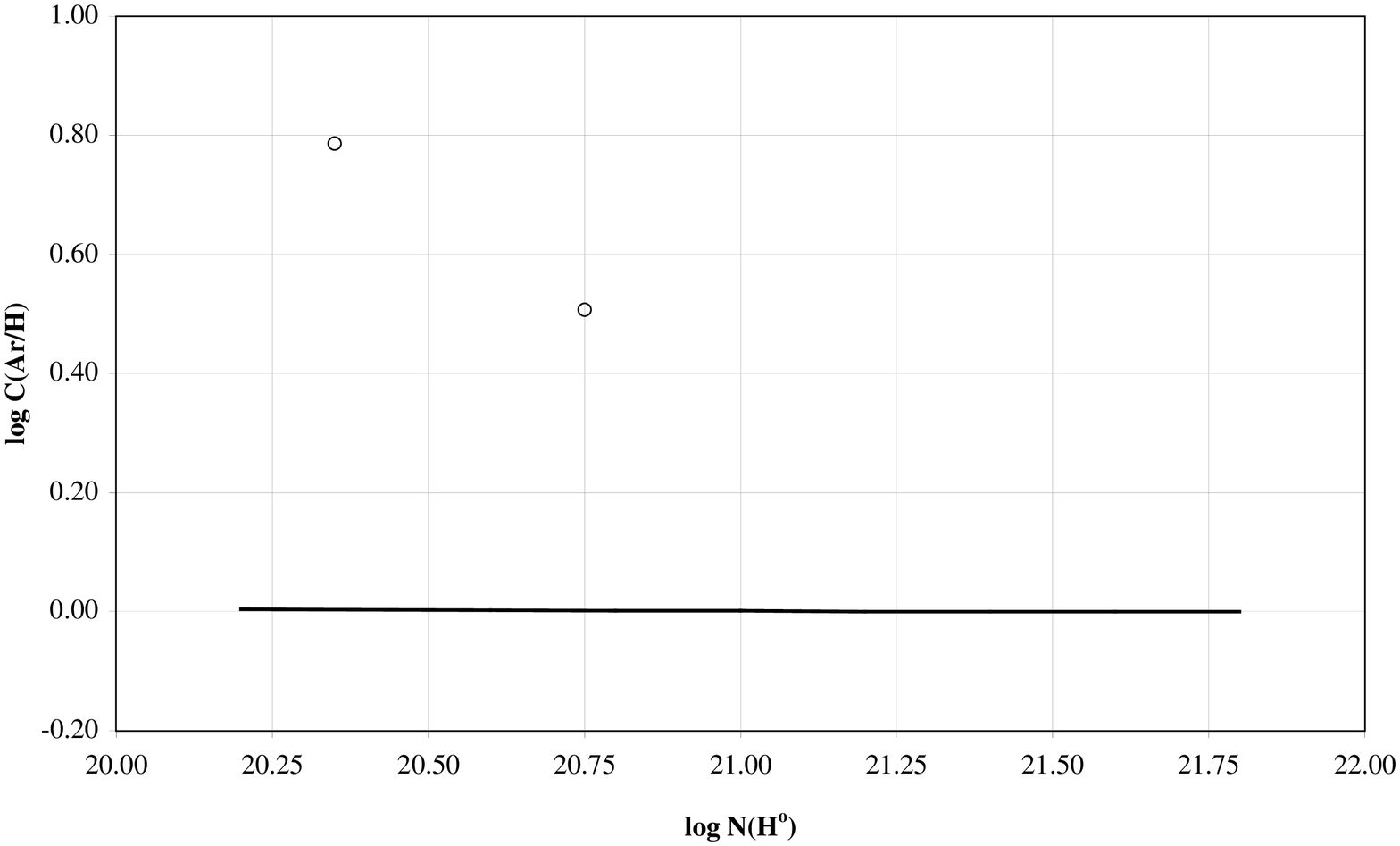}
\caption{
Ionization correction terms for  [Ar/H] measurements.
Legend as in Fig. \ref{cMgH}.} 
\label{cArH}
\end{figure}

\begin{figure}
\plotone{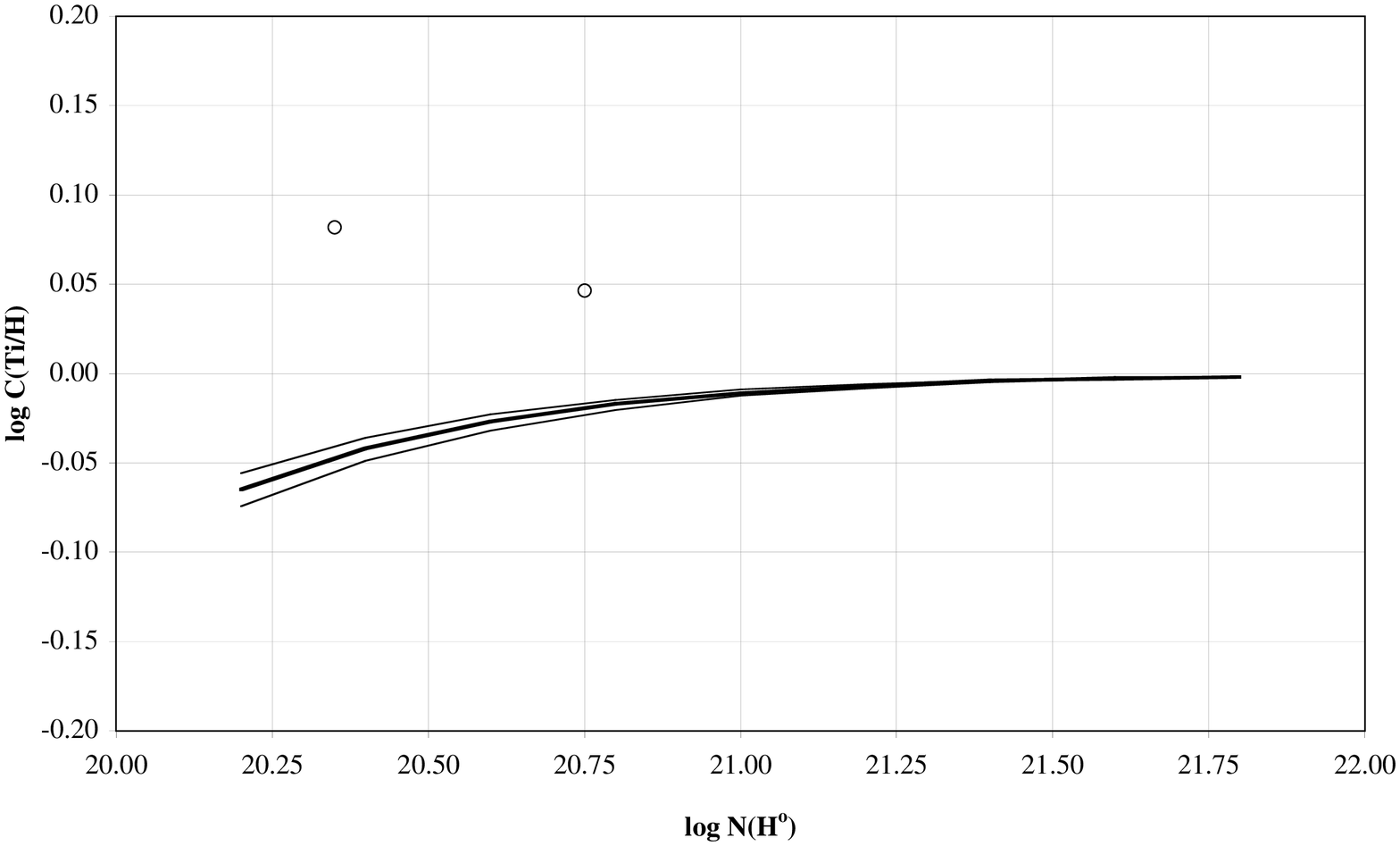}
\caption{
Ionization correction terms for  [Ti/H] measurements.
Legend as in Fig. \ref{cMgH}.} 
\label{cTiH}
\end{figure}

\begin{figure}
\plotone{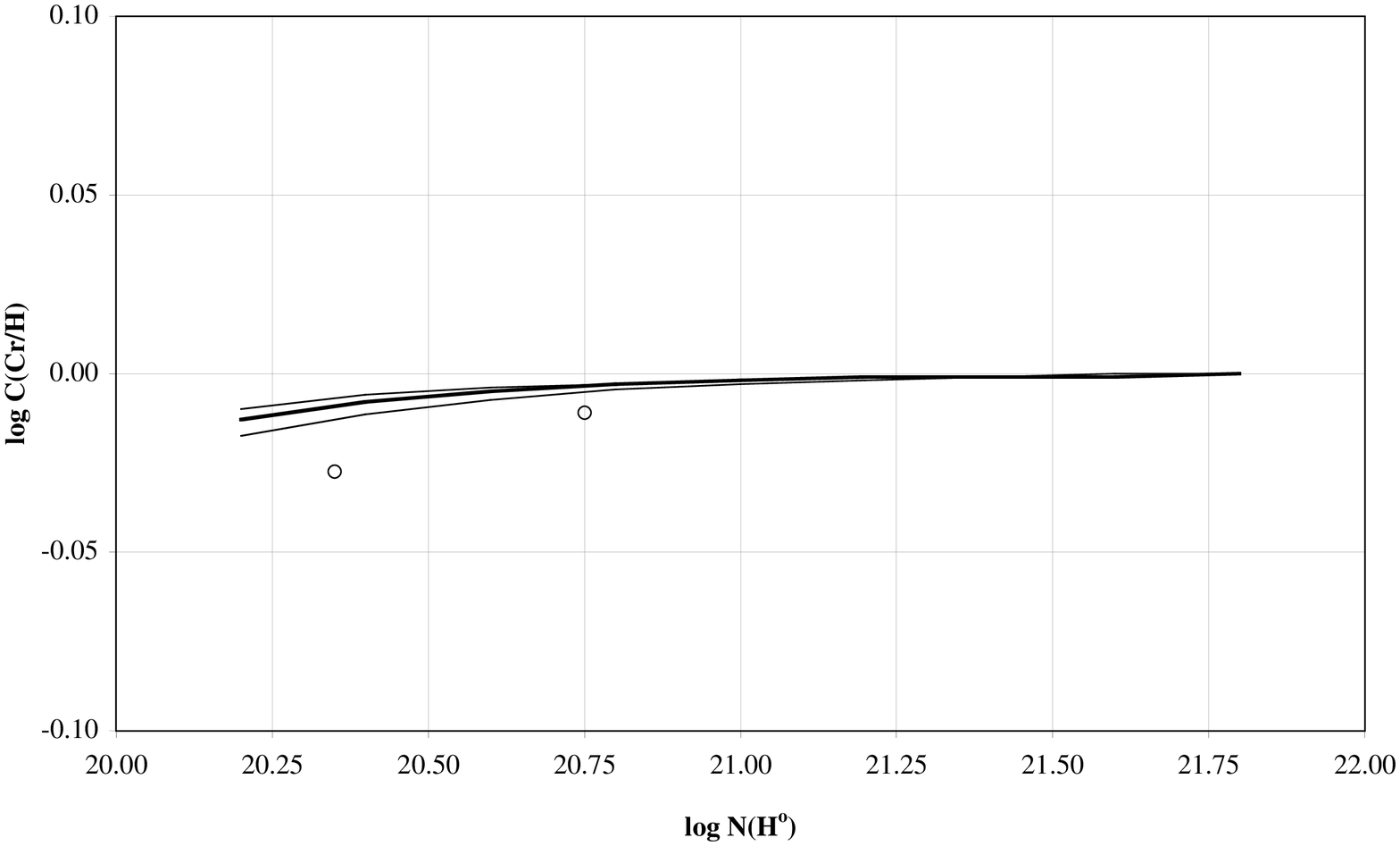}
\caption{
Ionization correction terms for  [Cr/H] measurements.
Legend as in Fig. \ref{cMgH}. } 
\label{cCrH}
\end{figure}

\begin{figure}
\plotone{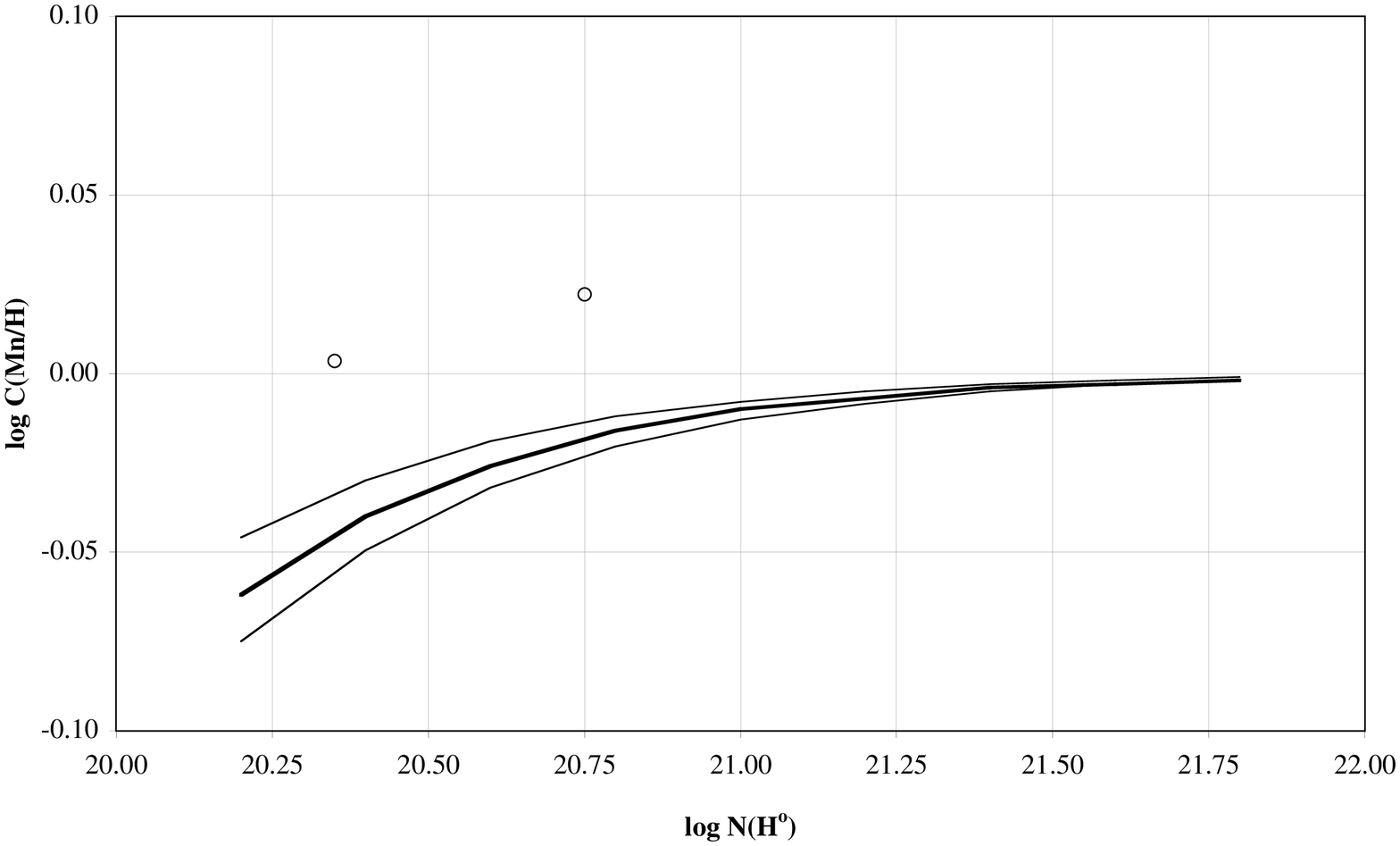}
\caption{
Ionization correction terms for  [Mn/H] measurements.
Legend as in Fig. \ref{cMgH}.}
\label{cMnH}
\end{figure}

\begin{figure}
\plotone{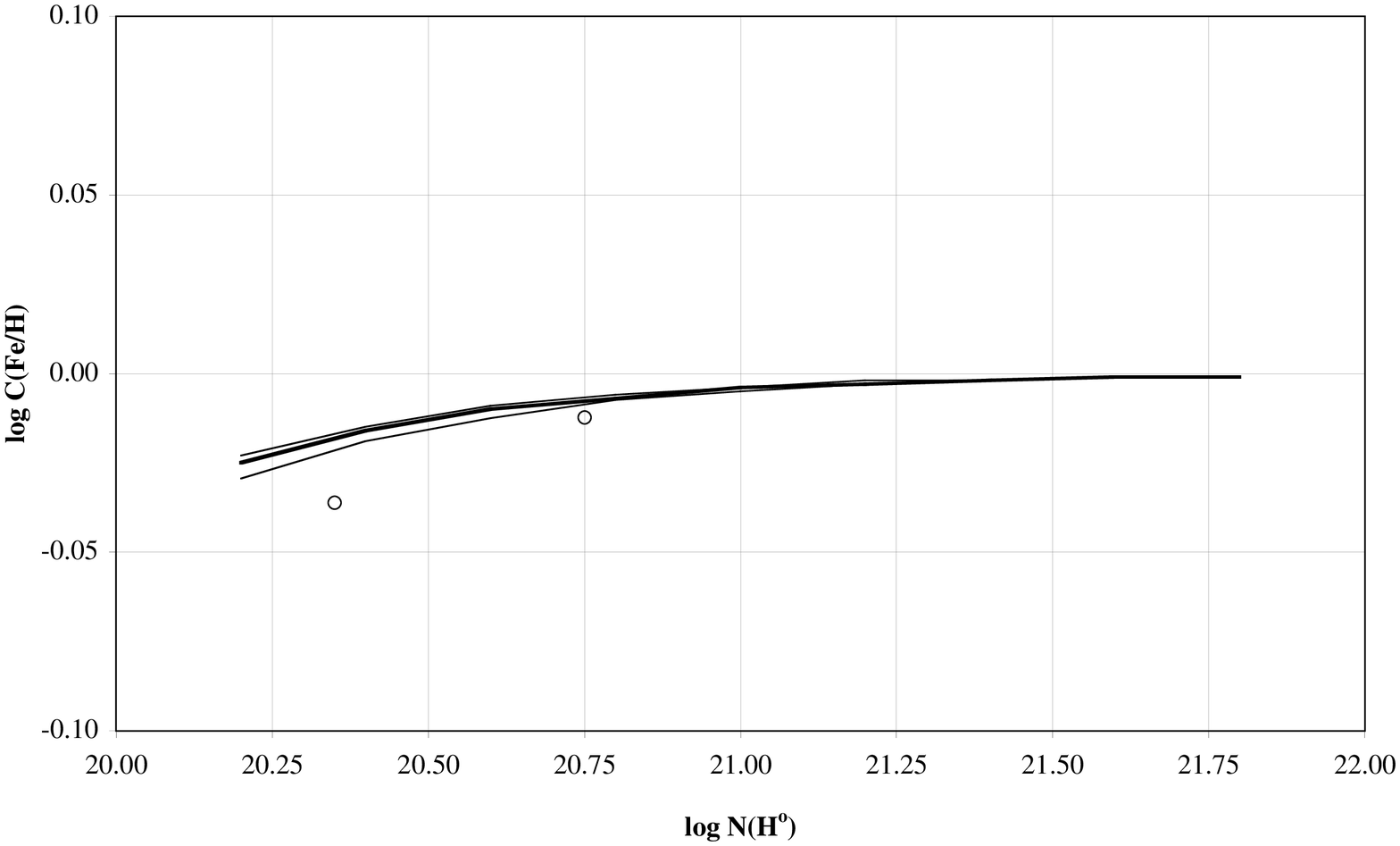}
\caption{
Ionization correction terms for  [Fe/H] measurements.
Legend as in Fig. \ref{cMgH}. } 
\label{cFeH}
\end{figure}

\begin{figure}
\plotone{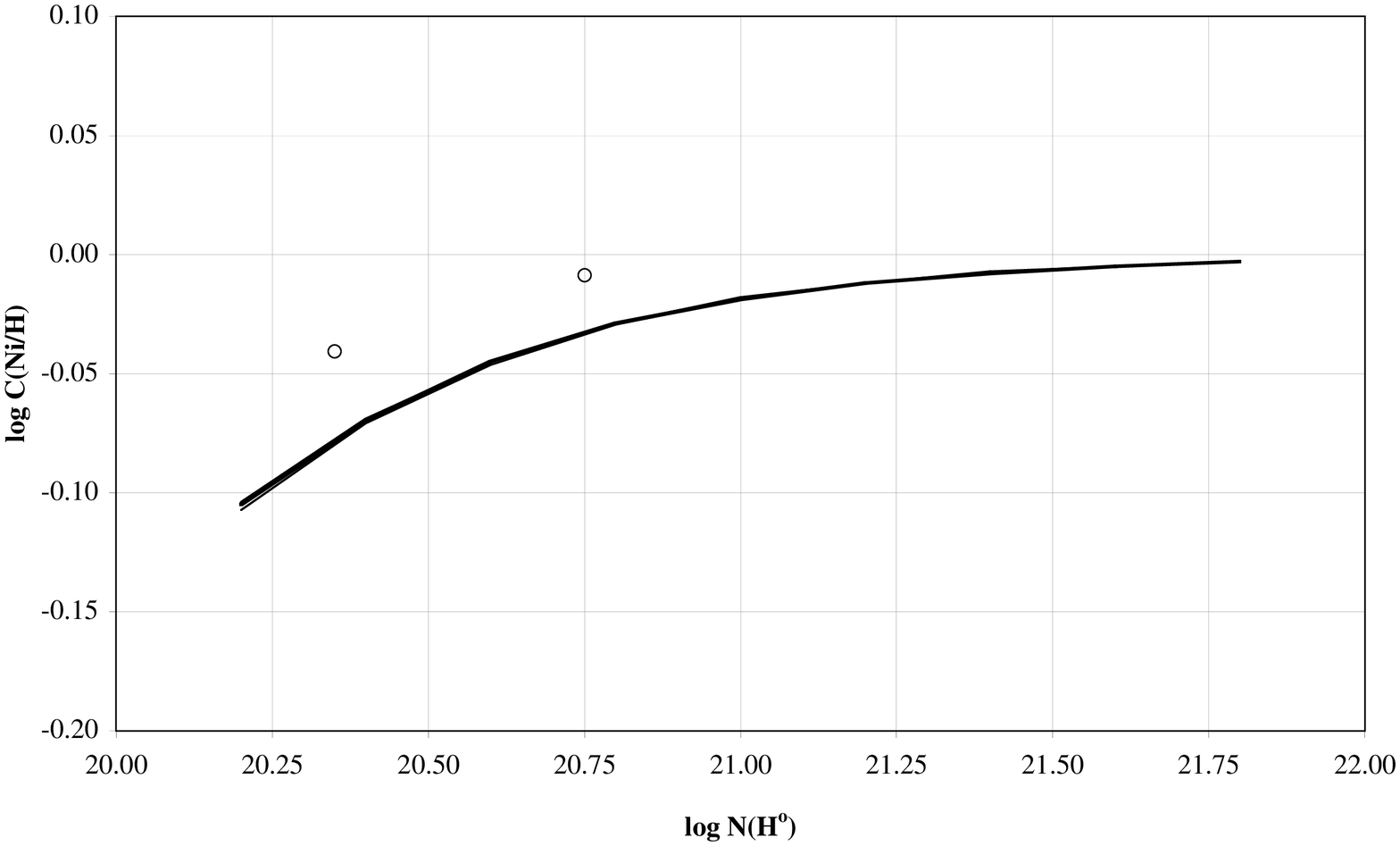}
\caption{
Ionization correction terms for  [Ni/H] measurements.
Legend as in Fig. \ref{cMgH}.} 
\label{cNiH}
\end{figure}

\begin{figure}
\plotone{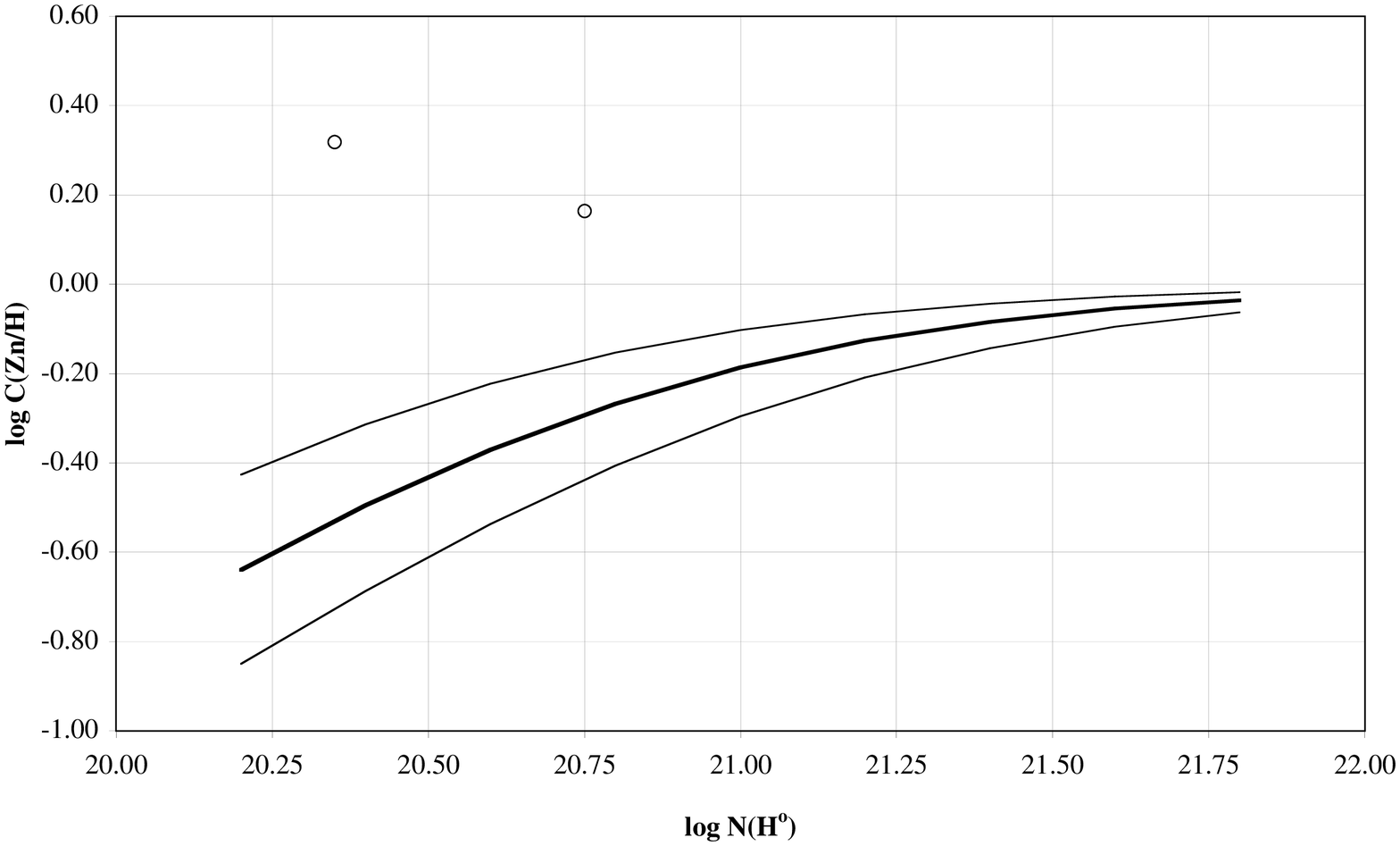}
\caption{
Ionization correction terms for  [Zn/H] measurements.
Legend as in Fig. \ref{cMgH}.} 
\label{cZnH}
\end{figure}

Abundance determinations in DLAs are based on column-density measurements of species which are
dominant ionization states in H$^{\circ}$ regions. 
In the Appendix we define
the ionization correction terms that allow us to recover the intrinsic
abundances starting from this type of measurements. 
In order to estimate
these   correction terms   we used the
same sets of input parameters --- ionizing spectrum,  $U$ and $N_1/N_2$ ---
that allow us to reproduce the observed 
\{$N$(H$^{\circ}$),${\cal R}$(Al$^{2+}$/Al$^{+}$)\} distribution. 
In practice,  with such parameters  
we determine the mean ionization fractions $\overline{x}_2({\rm X}^i)$
with CLOUDY
and   the correction terms from  Eqs. \ref{CXY2} and \ref{CXH2}.
Thanks to this procedure  the input parameters
 are   constrained by the requirement to model   the Al$^{2+}$/Al$^{+}$
observations.  
An additional advantage is that   the   ionization effects are estimated
as a function of $N$(H$^{\circ}$), which is a measurable quantity. 
In this way, the range of predicted ionization corrections  
is reduced  once we know
$N$(H$^{\circ}$)  for   individual DLAs. 
The resulting
  corrections for absolute abundances are shown in Figs. \ref{cMgH} through \ref{cZnH}. 
In each figure we display the   corrections  predicted by the S2 and H1 models discussed 
above. 
Corrections for N/H and O/H are not shown in the figures because they are always
below abundance measurement errors ($\lsim 0.05$ dex). 

The corrections estimated with  
 the S2 model  
are shown  
in Figs. \ref{cMgH} through \ref{cZnH} (absolute abundances) and Table 3
(relative abundances).   
In the S2 model
the species used for abundance measurements   arise mostly in type-1 neutral regions,
with an additional contribution from type-2 ionized regions. 
It is this latter contribution  that affects the 
measured abundances.
The correction terms predicted by this model are generally small owing to  
the high $U$ values  that we require
to match the ${\cal R}$(Al$^{2+}$/Al$^{+}$) versus $N$(H$^{\circ}$) data. 
In fact, when     $U$ is high, 
the species used for abundance measurements   tend  to vanish  in the ionized region
since they move to a higher state of  ionization.  
Ionization corrections tend to decrease in absolute value with increasing  $N$(H$^{\circ}$).
This is due to the fact  
that the relative contribution of the ionized layer becomes
less important at high $N$(H$^{\circ}$).

The predictions for the H1 model have been derived by considering
only the   solutions   which match   
the empirical 
$\log N$(H$^{\circ}$) - $\log {\cal R}$(Al$^{2+}$/Al$^{+}$) anticorrelation. 
The empty circles in Figs. \ref{cMgH} through \ref{cZnH} have been calculated
using the solutions found at the intersection between the linear regression
to the data points and the curves at constant ionization parameter
$\log U=-4.2$ and $-4.8$ shown in Fig. \ref{iAlHIQSOa}.
Owing to the difficulty of finding solutions at   lower $U$ and high $N$(H$^{\circ}$)
we have not calculated corrections at log\,$N$(H$^{\circ}$) $\geq 21$. 
It is clear, however, that the correction terms become negligible at high log\,$N$(H$^{\circ}$).
The reason for this is  that 
the corrections decrease (in absolute value) with decreasing
$U$ and, at the same time, $U \propto N({\rm H}^{\circ})^{-1.5}$. 
 The predicted corrections  are generally small because the match with
the observed ${\cal R}$(Al$^{+2}$/Al$^{+}$) ratios is found at low values of $U$.

The correction terms can be negative or positive depending on the 
model adopted and on the species considered. 
The corrections for absolute abundances X/H are negative in
the S2 model. In this model
the dominant species used for the   measurements,  
X$^{i_d}$ and H$^{\circ}$,
mostly arises in the type-1 region  unaffected by ionization;
the type-2 layer gives an extra contribution  
which enhances the  X$^{i_d}$/H$^{\circ}$ ratio and 
a negative correction is required to recover the intrinsic abundance.
In the   H1 model the corrections for absolute abundances can   be negative
or positive since  all the species  arise    in a single 
layer in which the  X$^{i_d}$/H$^{\circ}$ ratio can be larger or smaller than the intrinsic
abundance.

\subsection{ Accuracy of Al atomic data }

The above calculations rely on the accuracy of the input atomic data.
Here we are particularly interested in the accuracy
of Al   data since our models are constrained by the capability of matching
the observed $\log N$(H$^{\circ}$)-$\log {\cal R}$(Al$^{2+}$/Al$^{+}$)
anticorrelation.  
The  photoionization cross-sections currently used in CLOUDY calculations
are generally accurate within $\simeq$ 10\%, 
 the Al$^{+}$   cross section
having a regular level of accuracy  
(Verner et al. 1996; Ferland et al. 1998).   
Radiative recombination rate coefficients can be obtained with an accuracy
better than $\simeq 15$\% (Ferland et al. 1998). However, the recombination 
process can be dominated by dielectronic recombination (DR), which is a far more
uncertain mechanism.  
Low-temperature DR rates, which are critical
for determining the ionization balance in photoionization equilibrium, 
are lacking for many elements (Ferland et al. 1998).
Luckily, such coefficients have been calculated
for  Al.  However, 
given the theoretical and experimental uncertainties 
it is possible that  the Al$^{+}$ DR rate may be overestimated
(Nussbaumer \& Storey 1986).  
If the Al$^{+}$ recombination rate is overestimated, 
the predicted  ${\cal R}$(Al$^{2+}$/Al$^{+}$) ratio  
is  underestimated. 
An effect of this type has been reported in a photoionization   study
of a Lyman Limit system, in which  the   models that give a
good fit to other species are not able to reproduce the relatively high 
${\cal R}$(Al$^{2+}$/Al$^{+}$) ratio observed   
(D'Odorico \& Petitjean 2001).
In order to test the   consequences of an effect of this type we artificially
increased the ratio ${\cal R}$(Al$^{2+}$/Al$^{+}$)   
calculated  at any given value of  $U$.
The results that we found can be summarized as follows. 
In the S2 model  
(i) the observed anticorrelation is matched  at  lower    $U$  values;
(ii) the column density $N_2$ of the ionized layer, which scales with $U$, becomes lower;
(iii) the ionization corrections, which scale with $N_2$, become lower;
(iv) the Al correction becomes even lower (in absolute value)
owing to the reduced contribution of Al$^{+}$ from the ionized layer.  
In the H1 model
(i) the solutions that match the observed anticorrelation are shifted to lower $U$ 
values;
(ii) the ionization corrections, which in this case scale with $U$, become lower. 
In summary,  if the Al recombination rate is too high, then the
ionization parameter $U$ and the abundance corrections
calculated above should be reduced;
in particular,  
the Al abundance corrections of model S2 would be more in line
with those of the other elements. 

\subsection{ Implications for the indirect estimates of $N$(Al$^{+}$) }

The large values of Al correction terms that we find are somewhat surprising given the fact
that most Al$^{+}$ column densities have been indirectly estimated from  
Si$^{+}$  column densities.\footnote
{This discussion could be equally applied  
to the indirect estimates based on Fe$^{+}$ column densities.}
In fact,  the
$\log N$(Al$^{+}$)-$\log N$(Si$^{+}$)   correlation could, in principle, be destroyed
by the large ionization effects predicted. 
Since this is not the case,
we must understand why.  
One possible reason is that, owing to the
uncertainty of Al$^{+}$ recombination coefficients, the Al correction terms may
be overestimated, as we have discussed above. 
Another possibility is that the  Al correction terms are correct, 
but the correlation is not destroyed because we are considering a limited interval
of $N$(H$^{\circ}$).   
In fact, 
in the   column-density range
of the DLAs used for deriving the $\log N$(Al$^{+}$)-$\log N$(Si$^{+}$) correlation
--- i.e., 20.3   $<$ $\log N$(H$^{\circ}$) $<$  20.7 ---
the Al/Si  corrections terms  are   large, but nevertheless they only show 
 a modest variation
($\approx \pm 0.1$ dex).  
As a consequence, the   ionization effects  may significantly change 
the intercept
of the $\log N$(Al$^{+}$)-$\log N$(Si$^{+}$)   correlation,
but should not change significantly the dispersion and   the
slope.

The $N$(Al$^{+}$) values inferred from   $N$(Si$^{+}$)  
may contain a systematic error when $\log N$(H$^{\circ}$) $>$ 20.7
owing to the variation of  $\cal C$(Al/Si) with
$N$(H$^{\circ}$). If the Al correction terms are overestimated, the effect
is probably small. 
%
Otherwise, we
can quantify this error from the predicted variation of   $\log \cal C$(Al/Si)
between $\log N$(H$^{\circ}$) $\simeq 20.5$ and 
 $\log N$(H$^{\circ}$) $\simeq 21.5$.  
We find that $N$(Al$^{+}$) may be overestimated
by $\simeq 0.2/0.3$ dex in model S2
or underestimated by a similar  amount in  model H1.  
These values are generally within the error bars quoted in the last
column of Table 1.  
In any case, these errors would not affect  the results
of the present work. In the case of model S2, in order
to compensate for such errors we should decrease $N$(Al$^{+}$) 
and hence increase ${\cal R}$(Al$^{2+}$/Al$^{+}$) by $\simeq 0.2/0.3$ dex
at $\log N$(H$^{\circ}$) $>$  20.7. 
One can see from Fig. \ref{iAlHImodA} that such
increase  would even
improve the agreement between the S2 model   and the observations.  
In the case of the H1 model, we should instead decrease
${\cal R}$(Al$^{2+}$/Al$^{+}$). This would imply 
that the interval ($U_{\rm min}$,$U_{\rm max}$) of allowed
solutions should be slightly shifted to
lower $U$ values at high  $\log N$(H$^{\circ}$).

\section{ Implications for abundance studies }

We briefly discuss how  the ionization correction terms
presented here can affect  studies of DLA abundances. 
Unless differently specified, the results summarized below are relative
to both types of models considered in this work.

The   correction terms for N and O  are generally negligible. 
This is not surprising since the ionization fractions of H$^{\circ}$, N$^{\circ}$ and O$^{\circ}$ 
are held together because of the strong charge exchange reactions (e.g. Sembach et al. 2000).
We note, however, that  cosmic rays ionization can affect the 
N$^{\circ}$/O$^{\circ}$ ratio (Viegas 1995).
The [N/Si] ratio is slightly  underestimated (Table 3). 
The correction term for [N/S] can be positive or negative, depending
on the adopted model. In any case the effect is generally small, with
$|\log \cal C$(N/S)$|$ $\lsim +0.2$ dex for  $N$(H$^{\circ}$)  $\geq 20.3$. 
These results indicate that the large scatter of [N/Si] and [N/S] abundances found by 
Lu, Sargent \& (1998) and by Centuri\'on et al. (1998) is  
a genuine nucleosynthetic effect.  
Claims that nitrogen abundances are severely affected by ionization effects
(Izotov \& Thuan 1999) are not supported by our study.

The correction terms for Al may be quite large, but with   different 
signs depending on the model adopted (Fig. \ref{cAlH}).  
Measurements of the Al/Fe and Al/Si ratios in DLAs yield 
[Al/Fe] $\approx 0$, with  a scatter of $\simeq 0.3$ dex,  
and  [Al/Si] $\approx -0.4$ dex, with values   between $-0.1$ and $-0.6$ dex
(Prochaska \& Wolfe 1999).  
Local ISM studies indicate that Al 
and Fe have similar  depletion 
(Barker et al. 1984; Howk, Savage \& Fabian 1999).
The [Al/Fe] ratios should not be affected by depletion, while
the  [Al/Si] ones should be affected as much as the  
[Fe/Si] ratios, which are underestimated by $\approx 0.3$ dex in DLAs (Paper I).
So  we expect [Al/Fe] $\approx$ 0 
and [Al/Si] $\approx -0.1$ dex   after correcting for dust.
For the DLAs with available Al$^{+}$ measurements we predict
$\log \cal C$(Al/Fe) $\approx$
 $\log \cal C$(Al/Si) $\approx -0.4$ dex (S2 model) or
$\approx +0.2$ dex
(H1 model).
Therefore, after dust and ionization corrections 
are applied, the [Al/Fe] and [Al/Si] ratios
are below solar in the S2 model,
but somewhat enhanced  
in the H1 model. 
  As we mentioned above, these ionization effects may be weaker
  if the Al$^{+}$
recombination rate is overestimated.

Silicon and sulphur are  used as  a tracers of  alpha elements 
given the difficulty of measuring oxygen  in DLAs. 
The [Si/H] correction term  is $\simeq -0.1$ dex 
at $\log N$(H$^{\circ}$) $\simeq 20.3$ (Fig. \ref{cSiH})
and silicon abundances can be accordingly overestimated.
However, the effect is not strong and this may explain why  
[Si/H]  measurements in DLAs do not show a trend with ${\cal R}$(Al$^{2+}$/Al$^{+}$),  
as discussed in Section 2.3.  
The  [S/H] corrections terms are negative for the S2 model
and positive for the H1 model, in both cases being   
$\lsim 0.2$ dex in absolute value  (Fig. \ref{cSH}).

The [Si/Fe] ratio, which is   used as a proxy of the $\alpha$/Fe ratio,
 shows an enhancement of
$\simeq +0.3/+0.5$ dex in DLAs which has been interpreted
as an intrinsic nucleosynthetic effect (e.g. Lu et al. 1996) or as a differential dust depletion 
(e.g. Paper I). 
The present study indicates that 
[Si/Fe] is almost unaffected by ionization corrections, being possibly
overestimated by $\simeq$ 0.1 dex at low $N$(H$^{\circ}$) in the S2 model. 
Owing to the negligible dust depletion of both sulphur and zinc, the [S/Zn] ratio
has been used as a dust-free [$\alpha$/Fe] indicator in DLAs 
(Centuri\'on et al. 2000; see however Prochaska et al. 2000).  
The correction terms for the [S/Zn] ratio  are  $\approx  0.2$ dex in absolute value
for the DLAs with available S measurements. 
Considering that such corrections are dominated by the zinc contribution which may be
overestimated (see below), we conclude that the [S/Zn] results presented by Centuri\'on et al.( 2000)
are modestly affected by ionization effects.
Local interstellar studies yield a S/Zn ratio approximately solar
(Howk, Savage \& Sembach 1999) suggesting that ionization corrections are
unimportant.

Ionization effects tend to   lower the measured
Ar abundances in the local interstellar medium,
at least in lines of sight with $\log N$(H$^{\circ}$) $\leq 20.0$ (Sofia \& Jenkins 1998,
Jenkins et al. 2000).
In DLAs the Ar corrections are negligible in the case of the S2 model; however,
Ar may be severely underestimated  in the H1 model  when the H$^{\circ}$
column density is low (Fig. \ref{cArH}).

Corrections for Cr and Fe are $\lsim 0.05$ dex in absolute value
(Figs. \ref{cCrH} and \ref{cFeH})
and those for   Ti, Mn,   and Ni  are  
$\lsim 0.1$ dex 
(Fig. \ref{cTiH}, \ref{cMnH},   and \ref{cNiH}). 
Therefore, deviations from solar ratios observed for pairs of iron-peak
elements such as [Cr/Fe] and [Mn/Fe] (Lu et al. 1996; Prochaska \& Wolfe 1999)
must be due  either to dust depletion or to nucleosynthetic effects
since ionization effects are excluded.

The predicted corrections for Zn are relatively large, with opposite signs
depending on the   model adopted (Fig.  \ref{cZnH}). 
Owing to the predicted variation of the Zn correction terms with $\log N$(H$^{\circ}$)
we would expect  to find some trend between the [Cr/Zn] and [Fe/Zn] ratios
and $\log N$(H$^{\circ}$).   
The lack of any trend (see Section 2.4) suggests that the zinc correction terms may be
overestimated. 
This could be the case since 
the Zn recombination coefficients and   ionization cross sections  
are rather uncertain (Howk, Savage \& Fabian 1999).

\section{Summary and conclusions}

We have analyzed column density measurements of
Al$^{2+}$ and of singly-ionized species available
in literature to cast light on the   properties of
low-ionization regions in DLA systems.
We have found that $\log N$(Al$^{+}$) is   well correlated with 
$\log N$(Si$^{+}$) and  $\log N$(Fe$^{+}$)
and we  have used   this result to estimate $N$(Al$^{+}$) and the 
ratio ${\cal R}$(Al$^{2+}$/Al$^{+}$) $=$ $N$(Al$^{2+}$)/$N$(Al$^{+}$) for a sample
of 20 DLAs. 
The ratio can attain
relatively high values, up to $\simeq 0.6$, with   a median value of 0.2.
This result  is contrary to the common belief that the fraction
of  Al$^{2+}$  is generally small in DLAs. 
In the redshift interval $2.0 \leq z_{\rm abs} \leq 2.5$, where most of the
measurements are concentrated,  the ratio shows the full spread of a factor of 20.
Therefore, the ratio must be influenced by local effects  not dependent on    $z_{\rm abs}$.
Local absorption and/or local radiation fields  probably play an
important role in determining the ionization properties inside DLAs.

The presence of high fractions of Al$^{2+}$   with same velocity distribution of 
low-ionization  may
question the reliability of abundance determinations in DLAs.
We have  investigated the behavior of  [Si/H] and [Si/Fe] abundance ratios
in order to put in evidence
possible effects of ionization. However, we do not   find any trend
between these abundance ratios and  ${\cal R}$(Al$^{2+}$/Al$^{+}$).   
We have considered the possibility that
Al$^{2+}$ originates in a region 
different from the one where species of low ionization arise. 
In this case, the ratio ${\cal R}$(Al$^{2+}$/Al$^{+}$) would measure
the relative contribution of different  regions, rather than the
intrinsic degree of ionization of a single region. 
From a   regression analysis  of the logarithmic column densities
we find some evidence for a distinct behavior of
Al$^{2+}$. In fact, while  any pair of species including Al$^{+}$, Si$^{+}$ and Fe$^{+}$
yields correlations with slope unity and low dispersion, pairs including
Al$^{2+}$ yield correlations with   lower slopes     larger
dispersions.  One possible explanation of this observational result is that,
indeed, Al$^{2+}$ and singly-ionized species originate in two different regions. 
However, the two regions must be physically connected  
since the velocity profiles
of Al$^{2+}$ and of low ions are very similar. 
 
We have also identified the existence of an anticorrelation between  
$\log {\cal R}$(Al$^{2+}$/Al$^{+}$) and $\log N$(H$^{\circ}$). 
The anticorrelation appears to be an intrinsic property of DLAs
not induced by observational bias, at least as far as the
detection limits of Al$^{+}$ and Al$^{2+}$ are concerned. 
We have used such anticorrelation, in conjunction with photoionization equilibrium calculations,
to constrain the ionization parameter in DLAs and hence the
 abundance ionization corrections. 
We have proposed  that low-ionization species in DLAs may arise
in two types of regions:  (1) an H$^{\circ}$
region opaque to $h \nu > 13.6$ photons and/or   (2) a partially ionized,
Al$^{2+}$-bearing interface with small/negligible fractions of high ions such as Si$^{+3}$. 
We have considered two types of  ionizing continuum:
a soft, stellar-type  
($T_{\rm eff}$=33,000 K) and a hard, QSO-dominated type at $z \approx 2$.

We have succesfully reproduced the
observed    $\log {\cal R}$(Al$^{2+}$/Al$^{+}$) -log $N$(H$^{\circ}$)  
anticorrelation  by means of a soft-continuum, two-region  model (S2 model) 
with ionization parameter in the range 
10$^{-2.6}$ $\lsim U \lsim$ 10$^{-1.7}$.  
In this model 
most of the neutral hydrogen and low ionization species originates in the 
neutral region of type 1.
However, the total hydrogen column density of the  
partially ionized, type-2 region is relatively high,
$ 3  \times 10^{20} \, {\rm cm}^{-2}  \lsim  N_{2}({\rm H})  
\lsim 2  \times 10^{21} \, {\rm cm}^{-2}$.
At a given value of $U$, $N_{2}({\rm H})$ is fixed by the photoionization calculation
and the contribution of the type-2 region decreases as $N$(H$^{\circ}$) increases.
Because Al$^{2+}$ originates only in the type-2 region, the anticorrelation is
naturally explained. 

We also tried to reproduce the
observed    $\log {\cal R}$(Al$^{2+}$/Al$^{+}$) -log $N$(H$^{\circ}$)  
anticorrelation  by means of a hard-continuum, one-region model (H1 model). 
In this model all the species of low and intermediate ionization state are assumed
to arise in a single layer.  
In order to reproduce the   decrease of  $\log {\cal R}$(Al$^{2+}$/Al$^{+}$)
with the correct slope it is necessary to assume that  
$U$ must decrease  
with a law of the type
$U \propto N({\rm H}^{\circ}) ^{-1.5}$.
The anticorrelation might be due to the decrease of the mean
ionization level with increasing self-shielding by neutral hydrogen.  
The typical values of the ionization parameter at $\log {\rm N}({\rm H}^{\circ}) \simeq 20.8$ 
is $\log U \simeq -4.8$.

We have estimated abundance ionization corrections for 14 elements of astrophysical
interest both with the S2 and the H1 model. 
In both cases we used 
the same sets of parameters that allow us to reproduce the  anticorrelation 
between $\log {\cal R}$(Al$^{2+}$/Al$^{+}$) and log $N$(H$^{\circ}$). 
Ionization corrections
can be negative or positive
depending on the  model adopted and on the species considered. In any case
ionization corrections tend to  become smaller in absolute value
as  $N$(H$^{\circ}$) increases. 

Ionization corrections are small in both models, but for different reasons.
In the S2 model corrections are small because the species used for abundance
measurements tend to shift to a higher ionization state in the type-2 region
owing to the high value
of $U$ that we find. 
In the H1 model corrections are small because the 
single region where   low-ionization species, together with Al$^{2+}$, are located 
has a low level of ionization.

The correction terms  for the absolute abundances of  N, O, Ti, Cr, Mn, Fe,
and Ni are generally   below measurement errors  ($\approx 0.05/0.1$ dex),
independently of the adopted ionizing spectrum. 
Therefore the deviations from solar ratio observed
in some pairs of iron-peak elements, such as the [Mn/Fe] or
[Cr/Fe] ratios (Lu et al. 1996, Pettini et al. 2000)
are not induced from ionization effects. 
The Ar/H correction term is negligible for the S2 model
 but may be
significant for the H1 model, in which case the measured Ar$^{\circ}$/H$^{\circ}$
value would underestimate the intrinsic Ar/H abundance. 

The ionization corrections for Mg, Si, P, and S can attain values  somewhat higher
than the measurement errors.
The
[N/Si] and [N/S]  ratios  are modestly  affected by 
 ionization effects. As a consequence, 
the considerable [N/S] and [N/Si] scatter observed  
at a given metallicity (Centuri\'on et al. 1998; Lu, Sargent \& Barlow 1998)
is a genuine characteristic of DLAs.
The   [Si/Fe] ratio, a typical indicator of the $\alpha$/Fe-peak ratio, 
may be overestimated by $\approx 0.1$ dex at low $N$(H$^{\circ}$)
if ionization corrections are not applied.   The [S/Zn] ratio 
might be more sensitive
to ionization effects, but the result is uncertain since Zn corrections 
are probably inaccurate.

The Al corrections can be relatively large.
They can be negative or positive, depending whether we adopt the S2 or the H1 model,
respectively.  
The   [Al/Fe] and [Al/Si] ratios corrected for dust and ionization effects are
below the solar value in the S2    model and somewhat enhanced in the H1 model.
These effects, however, are less marked if the Al recombination rate is overestimated.

The Zn corrections are apparently large, but these results may be
 inaccurate owing to the   uncertainties of   Zn atomic parameters.  
From an analysis of [Cr/Zn] and [Fe/Zn] data versus $\log N$(H$^{\circ}$) 
we have provided evidence 
that zinc ionization corrections are likely to be overestimated.  
Therefore studies of DLAs metallicity  based on [Zn/H] data
(Pettini et al. 1997; Vladilo et al. 2000) are unlikely to be significantly affected by
ionization effects.

We have investigated the stability of the above results 
in light of possible inaccuracies of Al atomic parameters.
In particular we have considered the possibility, consistent with available data,
that  the Al$^+$ dielectronic recombination rate
may be overestimated. 
We find that in this case the  
ionization parameter $U$ and the abundance corrections
would be lower   both in the S2 and in the H1 model.

The ionization corrections presented here are significantly smaller than the ones
predicted by Izotov, Schaerer \& Charbonnel (2000), who have also considered a two-region model
of DLA gas. However,  these authors 
assume that the neutral region has  much lower metallicity than the ionized one,
a strong assumption for which there is little observational support
(see Levshakov, Kegel \& Agafonova, 2000).   
With such assumption the metal absorptions originate essentially in the ionized region
and the predicted ionization effects are obviously more enhanced than in our case. 

Future studies of ionization properties in DLAs would benefit from measurements
of  ionic ratios other than Al$^{2+}$/Al$^{+}$.
 One possibility is the Fe$^{2+}$/Fe$^{+}$ given the presence of the 112.2 nm 
transition of Fe$^{2+}$ which
could be observed in selected cases. From the  point of view of the atomic data
it is important to better understand      the actual accuracy of
atomic parameters, and in particular those of Al and Zn.

\acknowledgements
JCH acknowledges support from NASA Long Term Space Astrophysics grant NAG5-3485 through the
Johns Hopkins University. 
We thank the referee for suggestions that have significantly improved the quality of this work.

\section{ Appendix: ionization corrections and ionization ratios in the two-region model}

We assume that low ionization species in DLAs arise in two type of regions:
(1) an H$^{\circ}$ region completely opaque to ionizing photons with $h\nu > 13.6$\,eV and 
(2) a mildly ionized region containing intermediate-ionization species such as Al$^{2+}$,
but not high ions such as C$^{+3}$ or Si$^{+3}$.
  The  observed column density of 
the $i$-th ionization state of the element X is given by the
relation 
\begin{equation}
\label{Relation1}
N ({\rm X}^i) = \sum_{k=1,2}
\int_k  x_k({\rm X}^i) A_k({\rm X}) n_k({\rm H}) d l  
\end{equation}
where    $k = 1$ and $2$ indicates the neutral and ionized region, respectively;
the   integrals are carried on along 
the   portions of line of sight   $l$ through the two regions; 
$n_{k}({\rm X})  = \sum_i n_{k}({\rm X}^{i})$ is the local density
(atoms cm$^{-3}$) of X summed over all the possible
ionization states $i$; 
$x_{k}$(X$^{i}$) = $n_{k}$(X$^{i}$)/ $n_{k}({\rm X})$
is the ionization fraction  of the $i$-th   state; 
$A_{k}$(X) = $n_{k}$(X)/$n_{k}$(H) is the absolute abundance 
of  X.

We assume that type-1 and type-2 regions have equal abundances 
in a given DLA system: 
$A_1({\rm X}) = A_2({\rm X}) = A({\rm X})$. 
For the dominant ionization state in the type-1 region, $i_d$, we have 
$x_1({\rm X}^{i_d}) = 1$ and, in particular,
$x_1$(H$^{\circ}$) = 1. 
From these assumptions and from (\ref{Relation1})
we can express the intrinsic abundance ratio of two elements X and Y
 in terms of the column density ratio of the dominant species
\begin{equation}
\label{RelativeAbundance1}
{ A({\rm X}) \over A({\rm Y}) }=
{ N ({\rm X}^{i_d}) \over N ({\rm Y}^{i_d}) } 
\times {\cal C}({\rm X/Y}) ~,
\end{equation}
where
\begin{equation}
\label{CXY}
{\cal C}({\rm X/Y}) =
{
N_1({\rm H})  +
\int_2  x_2({\rm Y}^{i_d}) n_2({\rm H}) d l 
\over
N_1({\rm H}) +
\int_2  x_2({\rm X}^{i_d}) n_2({\rm H}) d l  
}   ~~
\end{equation}
is, by definition, the ionization correction factor 
and $N_1({\rm H})=\int_1 n_1({\rm H}) dl$   the mean hydrogen density
in the H$^{\circ}$ region.  
It is easy to obtain similar a expression for the correction term of absolute
abundances, ${\cal C}({\rm X/H})$, by replacing ${\rm Y}^{i_d}$ with ${\rm H}^{\circ}$
in Eq. (\ref{CXY}).

We can also derive 
the column density ratio between    two    ionization states of a
given  element.
In particular we are interested in comparing the ionization state which
is dominant in the neutral region, $i_{\rm d}$, with higher ionization states, $i_{\rm h}$.  
 Since 
$x_1({\rm X}^{i_h}) = 0$ and
$x_1({\rm X}^{i_d}) = 1$  we obtain  from (\ref{Relation1})
\begin{equation}
\label{AluminiumRatio}
{\cal R}({{\rm X}^{i_h} \over {\rm X}^{i_d}})
\equiv
{ N ({\rm X}^{i_h}) \over N ({\rm X}^{i_d}) }
=  
{
\int_2  x_2({\rm X}^{i_h}) n_2({\rm H}) d l  
\over
N_1({\rm H})   +
\int_2  x_2({\rm X}^{i_d}) n_2({\rm H}) d l  
} 
 ~.
\end{equation}

We can derive simpler expressions by 
introducing the average ionization fractions along the line of sight,
$\overline{x}_2({\rm X}^{i})= \int_2 x_2({\rm X}^{i}) n_2({\rm H}) d l    / N_2({\rm H}) $. 
With this definition we obtain\begin{equation}
\label{CXY2}
{\cal C}({\rm X/Y}) =
{
  \overline{x}_2({\rm Y}^{i_d})   + N_1/N_2
\over
  \overline{x}_2({\rm X}^{i_d}) + N_1/N_2
} ~~,
\end{equation} 
\begin{equation}
\label{CXH2}
{\cal C}({\rm X/H}) =
{
  \overline{x}_2({\rm H}^{\circ})  + N_1/N_2
\over
  \overline{x}_2({\rm X}^{i_d})   + N_1/N_2
} ~~,
\end{equation}
and 
\begin{equation}
\label{AluminiumRatio2} 
{\cal R}({{\rm X}^{i_h} \over {\rm X}^{i_d}}) 
=  
{
  \overline{x}_2({\rm X}^{i_h}) 
\over
  \overline{x}_2({\rm X}^{i_d})   + N_1/N_2
}  ~,
\end{equation}
where $N_1/N_2 \equiv N_1({\rm H})/N_2({\rm H})$ is
the fraction of the line-of-sight total hydrogen column densities
in the two regions.
 The mean ionization fractions  $\overline{x}_2({\rm X}^{i})$
can be estimated by modeling the intensity and spectrum of
the  ionizing radiation field in which the clouds are embedded. 
The parameter $N_1/N_2$
plays a central role in assessing the importance of ionization effects.  
When $ { N_1  /  N_2 } \gg 1$
 the  ionization corrections are negligible since
 ${\cal C}({\rm X/Y}) \simeq 1$ 
and ${\cal C}({\rm X/H}) \simeq 1$.
In this case the ionization ratio ${\cal R}({\rm X}^{i_h}/ {\rm X}^{i_d})$
tends to be very low
no matter which are the conditions in the ionized envelope.
On the other hand, if ${ N_2 /  N_1 } \gg 1$,
the ionization corrections and the ionization ratio are representative
of the ionized envelope and not of the neutral region.
This latter case is appropriate when we consider that all species
of low and intermediate ionization state originate in a single,
partially ionized layer.

\newpage

\begin{deluxetable}{lccccccc}
\tabletypesize{\scriptsize}
\tablecolumns{8}
\tablenum{1}
\footnotesize
\tablecaption{Al$^{+}$ and Al$^{2+}$ column densities in DLA systems.}
\tablewidth{0pt}
\tablehead{
\colhead{QSO} & \colhead{$z_{\rm abs}$}   & 
\colhead{log $N_{\rm HI}$} & 
 \colhead{log $N$(Al$^{2+}$) } & \colhead{log $N$(Al$^{+}$) } &
\colhead{Ref}  &
\colhead{log $N$(Al$^{+}$)$_{\rm pred}$\tablenotemark{a} } &
\colhead{log ${ N ({\rm Al}^{2+} ) \over  N ({\rm Al}^{+} ) }$  }
} 
\startdata
0000$-$263 & 3.3901 & 21.41 $\pm$ 0.08  &  12.54 $\pm$ 0.10\tablenotemark{b} & $>$ 13.15& 2 & 13.66 $\pm$ 0.17& $-1.12\pm0.20$ \\ 
0100+130 & 2.3090 & 21.40 $\pm$ 0.05  & 12.72 $\pm$ 0.03 & ---& 1 & 14.17 $\pm$ 0.25 & $-1.45\pm0.25$\\
0149+33 & 2.1400  & 20.50 $\pm$ 0.10  & 12.56 $\pm$ 0.04 & 12.94 $\pm$ 0.10& 1  & 13.14 $\pm$ 0.18& $-0.38\pm0.11$ \\
0201+365	& 2.4620	& 20.38	$\pm$ 0.04	& 	13.61	$\pm$ 0.01	& -- & 5 &	 14.14 $\pm$ 0.24 & $-0.53 \pm  0.17$ \\ 
0216+080 &1.7688 & 20.00 $\pm$ 0.18  & 13.20 $\pm$ 0.07 & ---& 2 & 13.46 $\pm$ 0.20& $-0.26\pm0.21$\\
0216+080 & 2.2931 & 20.45 $\pm$ 0.16  & 13.74 $\pm$ 0.02 & $>$ 13.88& 2  & 14.03 $\pm$ 0.22& $-0.29\pm0.22$\\
0307$-$4945 & 4.466~ & 20.67 $\pm$ 0.09 & --- & 13.36 $\pm$ 0.06 & 7 & 13.28 $\pm$ 0.17 & --- \\
0458$-$02 & 2.0400 & 21.65 $\pm$ 0.09  & 13.34 $\pm$ 0.02 & $>$ 13.72& 1  & --- & --- \\
0528$-$2505 & 2.1410 & 20.70 $\pm$ 0.08  & 12.77 $\pm$ 0.08 & $>$ 13.46& 2  & 13.84 $\pm$ 0.19& $-1.07\pm0.20$\\
0528$-$2505 &  2.8110 & 21.20 $\pm$ 0.10 & $>$ 14.07 & $>$ 14.20 & 2  & ---& --- \\
0841+129 & 2.3745 & 20.95 $\pm$ 0.09  & 12.60 $\pm$ 0.10\tablenotemark{c}  & $>$ 13.26 & 1  & 13.82 $\pm$ 0.18 & $-1.21\pm0.21$\\
0841+129 & 2.4764 & 20.78 $\pm$ 0.10  & 12.63 $\pm$ 0.04 & $>$ 13.19\tablenotemark{d}& 1  & --- & --- \\
0951-0450 & 3.8570 & 20.60 $\pm$ 0.10 & --- & 13.30 $\pm$ 0.02& 1 & 13.22 $\pm$ 0.17 & ---\\
1104$-$1805 & 1.6616 & 20.85 $\pm$ 0.01  & 13.06 $\pm$ 0.02 & $>$13.43\tablenotemark{d,e}& 3  & 13.96 $\pm$ 0.21 & $-0.90\pm0.21$ \\
1215+33 & 1.9990 & 20.95 $\pm$ 0.07  & 12.78 $\pm$ 0.02 & $>$ 13.36& 1  & 13.61 $\pm$ 0.17& $-0.84\pm0.17$ \\
1331+170 & 1.7764 & 21.18 $\pm$ 0.04  & 12.97 $\pm$ 0.01\tablenotemark{f}& $>$ 13.57& 1  & 13.86 $\pm$ 0.19 & $-0.88\pm0.19$\\
1346$-$0322 & 3.7360 & 20.70 $\pm$ 0.10  & --- & 12.55 $\pm$ 0.03 & 1 & 12.52 $\pm$ 0.17 & --- \\
1425+603 & 2.8268 & 20.30 $\pm$ 0.04   & --- & $>$ 13.55& 2 & --- & ---\\
1795+75  & 2.6250 & 20.80 $\pm$ 0.10 & 13.62 $\pm$ 0.04\tablenotemark{f}& ---& 1 & 14.11 $\pm$ 0.24 & $-0.49\pm0.24$\\
1946+765 & 1.7382 & ---  & 12.70 $\pm$ 0.04 & ---& 2 & 13.33 $\pm$ 0.17 & $-0.63\pm0.18$\\
1946+765 & 2.8443 & 20.27 $\pm$ 0.06  & $< 12.09$ & 12.26 $\pm$ 0.03& 2,6 & 12.19 $\pm$ 0.17 & $<-0.17$ \\
2206$-$199A & 1.9200 & 20.65 $\pm$ 0.10  & 13.88 $\pm$ 0.01\tablenotemark{f}& --- & 4 & 14.39 $\pm$ 0.28 & $-0.51\pm0.28$\\
2206$-$199A & 2.0760 & 20.43 $\pm$ 0.10  & --- & 12.18 $\pm$ 0.01& 4 & 12.32 $\pm$ 0.17 & --- \\
2230+025 & 1.8642 &  20.85 $\pm$ 0.08 & 13.60 $\pm$ 0.01\tablenotemark{f} & $>$ 14.02 & 1  & 14.24 $\pm$ 0.26 & $-0.64\pm0.26$\\
2231$-$0015 & 2.0662 & 20.56 $\pm$ 0.10 & 13.14 $\pm$ 0.03\tablenotemark{c}  & --- & 1 & 13.83 $\pm$ 0.19 & $-0.69\pm0.19$\\
2237$-$0607 & 4.0803 & 20.52 $\pm$ 0.11 & --- & 12.85 $\pm$ 0.02 & 2 & 12.84 $\pm$ 0.17 & ---\\
2348$-$147 & 2.2790 & 20.56 $\pm$ 0.08 & 11.95 $\pm$ 0.08\tablenotemark{g}& 12.65 $\pm$ 0.01 & 1  & 12.79 $\pm$ 0.17 & $-0.70\pm0.08$\\
2359$-$0216 &  2.0950 & 20.70 $\pm$ 0.10 & 12.88 $\pm$ 0.05\tablenotemark{g}& $>$ 13.66\tablenotemark{d}   & 1  & 13.99 $\pm$ 0.23 & $-1.11\pm0.24$\\
2359$-$0216 & 2.1540 & 20.30 $\pm$ 0.10 & 12.96 $\pm$ 0.02 & 13.17 $\pm$ 0.02 & 1  & 12.88 $\pm$ 0.20 & $-0.20\pm0.03$\\
\enddata  
\tablenotetext{a}{ Value  estimated from $N$(Si$^{+}$) and 
from the linear relation  shown in Fig. 1.
The error bar is obtained from error propagation of the $N$(Si$^{+}$) error
and of  the linear regression error. The latter is estimated by taking into account
the dispersion of the linear regression, $\sigma$, and
the errors $\sigma_m$ and $\sigma_q$ given in  the third row of Table 2. } 
\tablenotetext{b}{Value determined from the UVES spectrum presented by Molaro et al. (2000a).
Only the Al$^{2+}$ line at 186.2 nm has been used. The line at 185.4 nm is 
affected by telluric contamination.}
\tablenotetext{d}{The published value  is treated as a lower limit since
the line is saturated in a substantial part of the profile.}
\tablenotetext{c}{Mean value of the lines at $\lambda$ 185.4 and 186.2 nm.}
\tablenotetext{e}{ The profile fitting result given by the authors 
is not reliable owing to the strong saturation of the line; we adopt as a lower limit
the result of the 
optical depth analysis.}
\tablenotetext{f}{ The Al$^{2+}$ absorption profile shows one or more
features
which are not seen, or are much weaker, in the absorption profiles of low ions; 
the equivalent width of such features
is significantly smaller than that of the bulk of the absorption  
seen also in low ions. 
}  
\tablenotetext{g}{
The column density refers only to the Al$^{2+}$ 
absorption profile in the  radial velocity range  where
the bulk of low ions is observed; 
a high-velocity Al$^{2+}$ absorption component is also present,
but its column density is not reported here.} 
\tablerefs{  (1) Prochaska \& Wolfe (1999); (2) Lu et al. (1996);
(3) Lopez et al. (1999); (4) Prochaska \& Wolfe (1997); 
(5) Prochaska \& Wolfe (1996);
(6) Lu et al. (1995); (7) Dessauges-Zavadsky et al. (2001) }
\end{deluxetable}

\newpage 
\clearpage
 
\begin{deluxetable}{crrrrr}
\tablecolumns{6}
\tablenum{2}
\footnotesize
\tablecaption{ Linear regression analysis of logarithmic column densities in DLA systems.}
\tablewidth{0pt}
\tablehead{
\colhead{ Ions }   & 
\colhead{$n$} & 
\colhead{$r$ } & \colhead{ $\sigma$ } & 
\colhead{$m \pm \sigma_m$}  &
\colhead{$q \pm \sigma_q$ }  
} 
\startdata  
Si$^{+}$ vs. Fe$^{+}$& 30 & 0.96 &0.17   &0.99$\pm$0.06   &0.64$\pm$0.82\\ 
Al$^{+}$ vs. Fe$^{+}$& 8 & 0.91 &0.20    &1.20$\pm$0.22    &-3.73$\pm$3.10\\
Al$^{+}$ vs. Si$^{+}$& 9 & 0.94 &0.16    &1.05$\pm$0.15    &-2.20$\pm$2.11\\ 
Al$^{2+}$ vs. Fe$^{+}$&20&0.51&0.37  &0.46$\pm$0.18  &6.25$\pm$2.70\\
Al$^{2+}$ vs. Si$^{+}$&17&0.74&0.35  &0.83$\pm$0.19  &0.45$\pm$2.94\\ 
\enddata  
\end{deluxetable}


\begin{deluxetable}{lccccccccc}
\tabletypesize{\scriptsize}
\tablecolumns{10}
\tablenum{3}
\footnotesize
\tablecaption{ Ionization correction terms for  abundance ratios   
in DLA systems\tablenotemark{a}}
\tablewidth{0pt}
\tablehead{
\colhead{ $ \log N$(H$^{\circ}$)} & 
\colhead{20.2} & \colhead{20.4} & \colhead{20.6} & \colhead{20.8} & 
\colhead{21.0} & \colhead{21.2} & \colhead{21.4} & \colhead{21.6} & \colhead{21.8}   
} 
\startdata 
$\cal C$[N/O] & +0.002	& +0.001	& +0.001	& +0.001	& +0.000	& +0.000	& +0.000	& +0.000	& +0.000 \\ 
$\cal C$[N/Si] & +0.149	& +0.100	& +0.066	& +0.043	& +0.028	& +0.018	& +0.011	& +0.007	& +0.004 \\ 
$\cal C$[N/S] & +0.227	& +0.157	& +0.105	& +0.070	& +0.045	& +0.029	& +0.019	& +0.012	& +0.007 \\ 
$\cal C$[N/Fe] & +0.027	& +0.017	& +0.011	& +0.007	& +0.004	& +0.003	& +0.002	& +0.001	& +0.001 \\ 
$\cal C$[N/Zn]\tablenotemark{b} & +0.642	& +0.498	& +0.372	& +0.269	& +0.188	& +0.128	& +0.085	& +0.055	& +0.036 \\ 
$\cal C$[O/Mg] & +0.152	& +0.102	& +0.068	& +0.044	& +0.028	& +0.018	& +0.011	& +0.007	& +0.005 \\ 
$\cal C$[O/Si] & +0.147	& +0.099	& +0.065	& +0.042	& +0.027	& +0.017	& +0.011	& +0.007	& +0.004 \\ 
$\cal C$[O/S] & +0.225	& +0.155	& +0.105	& +0.069	& +0.045	& +0.029	& +0.018	& +0.012	& +0.007 \\ 
$\cal C$[O/Fe] & +0.025	& +0.016	& +0.010	& +0.006	& +0.004	& +0.003	& +0.002	& +0.001	& +0.001 \\ 
$\cal C$[O/Zn]\tablenotemark{b} & +0.640	& +0.496	& +0.371	& +0.268	& +0.187	& +0.127	& +0.085	& +0.055	& +0.036 \\ 
$\cal C$[Mg/Si] & $-$0.005	& $-$0.004	& $-$0.002	& $-$0.002	& $-$0.001	& $-$0.001	 & $-$0.000	& $-$0.000	& $-$0.000 \\ 
$\cal C$[Mg/S] & +0.073	& +0.053	& +0.037	& +0.025	& +0.017	& +0.011	& +0.007	& +0.004	& +0.003 \\ 
$\cal C$[Mg/Fe] & $-$0.127	& $-$0.086	& $-$0.057	& $-$0.037	& $-$0.024	& $-$0.015	& $-$0.010	& $-$0.006	& $-$0.004 \\ 
$\cal C$[Mg/Zn]\tablenotemark{b} & +0.488	& +0.394	& +0.304	& +0.224	& +0.159	& +0.109	& +0.073	& +0.048	& +0.031 \\ 
$\cal C$[Al/Si] & $-$0.531	& $-$0.431	& $-$0.334	& $-$0.249	& $-$0.178	& $-$0.123	& $-$0.082	& $-$0.054	& $-$0.035 \\ 
$\cal C$[Al/Fe] & $-$0.653	& $-$0.514	& $-$0.389	& $-$0.284	& $-$0.201	& $-$0.137	& $-$0.092	& $-$0.060& $-$0.039 \\ 
$\cal C$[Al/Zn]\tablenotemark{b} & $-$0.038	& $-$0.033	& $-$0.028	& $-$0.023	& $-$0.017	& $-$0.013	& $-$0.009	& $-$0.006	& $-$0.004 \\ 
$\cal C$[Si/Fe] & $-$0.122	& $-$0.083	& $-$0.055	& $-$0.036	& $-$0.023	& $-$0.015	& $-$0.009	& $-$0.006	& $-$0.004 \\ 
$\cal C$[Si/Zn]\tablenotemark{b} & +0.493	& +0.397	& +0.306	& +0.226	& +0.160	& +0.110	& +0.074	& +0.048	& +0.031 \\ 
$\cal C$[P/Si] & $-$0.150	& $-$0.111	& $-$0.079	& $-$0.054	& $-$0.036	& $-$0.024	& $-$0.015	& $-$0.010	& $-$0.006 \\ 
$\cal C$[P/Fe] & $-$0.272	& $-$0.194	& $-$0.134	& $-$0.090& $-$0.059	& $-$0.039	& $-$0.025	& $-$0.016	& $-$0.010 \\ 
$\cal C$[P/Zn]\tablenotemark{b} & +0.343	& +0.286	& +0.227	& +0.172	& +0.124	& +0.086	& +0.058	& +0.038	& +0.025 \\ 
$\cal C$[S/Si] & $-$0.078	& $-$0.057	& $-$0.039	& $-$0.027	& $-$0.018	& $-$0.011	& $-$0.007	& $-$0.005	& $-$0.003 \\ 
$\cal C$[S/Fe] & $-$0.200	& $-$0.139	& $-$0.094	& $-$0.063	& $-$0.041	& $-$0.026	& $-$0.017	& $-$0.011	& $-$0.007 \\ 
$\cal C$[S/Zn]\tablenotemark{b} & +0.415	& +0.341	& +0.267	& +0.199	& +0.143	& +0.098	& +0.066	& +0.044	& +0.028 \\ 
$\cal C$[Ti/Si] & +0.083	& +0.057	& +0.038	& +0.025	& +0.016	& +0.010	& +0.007	& +0.004	& +0.003 \\ 
$\cal C$[Ti/Fe] & $-$0.039	& $-$0.026	& $-$0.017	& $-$0.011	& $-$0.007	& $-$0.004	& $-$0.003	& $-$0.002	& $-$0.001 \\ 
$\cal C$[Ti/Zn]\tablenotemark{b} & +0.576	& +0.454	& +0.344	& +0.251	& +0.176	& +0.120	& +0.08	& +0.052	& +0.034 \\ 
$\cal C$[Cr/Si] & +0.134	& +0.091	& +0.060& +0.039	& +0.025	& +0.016	& +0.010& +0.006	& +0.004 \\ 
$\cal C$[Cr/Fe] & +0.012	& +0.008	& +0.005	& +0.003	& +0.002	& +0.001	& +0.001	& +0.001	& +0.000 \\ 
$\cal C$[Cr/Zn]\tablenotemark{b} & +0.627	& +0.488	& +0.366	& +0.265	& +0.185	& +0.126	& +0.084	& +0.055	& +0.035 \\ 
$\cal C$[Mn/Si] & +0.086	& +0.059	& +0.039	& +0.026	& +0.017	& +0.011	& +0.007	& +0.004	& +0.003 \\ 
$\cal C$[Mn/Fe] & $-$0.036	& $-$0.024	& $-$0.016	& $-$0.01	& $-$0.006	& $-$0.004	& $-$0.003	& $-$0.002	& $-$0.001 \\ 
$\cal C$[Mn/Zn]\tablenotemark{b} & +0.579	& +0.456	& +0.346	& +0.252	& +0.177	& +0.121	& +0.080& +0.053	& +0.034 \\ 
$\cal C$[Ni/Si] & +0.042	& +0.029	& +0.020 & +0.013	& +0.009	& +0.005	& +0.004	& +0.002	& +0.001 \\ 
$\cal C$[Ni/Fe] & $-$0.080& $-$0.054	& $-$0.035	& $-$0.023	& $-$0.015	& $-$0.009	& $-$0.006	& $-$0.004	& $-$0.002 \\ 
$\cal C$[Ni/Zn]\tablenotemark{b} & +0.535	& +0.427	& +0.326	& +0.239	& +0.169	& +0.115	& +0.077	& +0.050& +0.033 \\ 
$\cal C$[Zn/Fe] & $-$0.615	& $-$0.480	& $-$0.361	& $-$0.262	& $-$0.183	& $-$0.125	& $-$0.083	& $-$0.054	& $-$0.035 \\ 
\enddata  
\tablenotetext{a}{\,Results predicted by model S2 at $\log U =-2.2$ (see Section 3.1).
All values are given in logarithm. }
\tablenotetext{b}{\,Systematic errors probably present owing to uncertainty of zinc
atomic parameters}
\end{deluxetable}

\newpage
\clearpage


\begin{references}


 
\reference{}
Barker, E.S., Lugger, P.M., Weiler, E.J., \& York, D.G. 1984, \apj, 280, 600

\reference{}
 Boiss\'e, P., Le Brun, V., Bergeron, J., \& Deharveng, J.M.
1998, \aap, 333, 841

\reference{} 
Centuri\'on M., Bonifacio P., Molaro P., \& Vladilo G. 1998,
\apj,  509, 620

 
\reference{} 
Centuri\'on M., Bonifacio P., Molaro P., \& Vladilo G. 2000,  \apj, 536, 540

 
\reference{} 
D'Odorico, V., \& Petitjean, P. 2001, \aap, in press (astro-ph/0102491)

\reference{}
Ferland, G.J., 1996, A Brief Introduction to CLOUDY 90
(Univ. of Kentucky, Dep. of Physics and Astronomy, Internal Report)

\reference{}
Ferland, G.J., Korista, K.T., Verner, D.A., Ferguson, J.W., Kinkgdon, J.B., \&
Verner, E.M. 1998, \pasp, 110, 761

 

\reference{} 
Haardt, F., \& Madau, P. 1996, \apj, 461, 20
 
\reference{} 
 Howk, J.C., Savage, B.D., \& Fabian, D. 1999, \apj,  525, 253

\reference{} 
 Howk, J.C., \& Sembach, K.R. 1999, \apj, 523, L141  (Paper II)

 
 
\reference{} 
Izotov, Y.I., Schaerer, D., \& Charbonnel, C., 2000,\apj, in press (astro-ph/0010643)


\reference{} 
Izotov, Y.I., \& Thuan, T.X. 1999, \apj, 511, 639

\reference{}
Jenkins, E.B., Oegerle, W.R., Gry, C., Vallerga, J., Sembach, K.R., Shelton, R.L., Ferlet, R.,
Vidal-Madjar, A., York, D.G., Linsky, J.L., Roth, K.C., Dupree, A.K., \& Edelstein, J.
2000, \apj, 538, L81

\reference{} 
Kulkarni, V.P., Fall, S.M., \& Truran, J.W., 1997, \apj, 484, L7

\reference{} 
Kurucz, R.L. 1991, in Proc. Workshop on Precision Photometry: Astrophysics of the Galaxy,
eds. A. C. Davis Philip, A. R. Upgren, \& K. A. James (Schenectady: Davis), 27

\reference{}
Lauroesch, J.T., Truran, J.W., Welty, D.E., \& York, D.G., 1996, \pasp, 108, 641

\reference{}
Levshakov, S.A., Kegel, W.H., \& Agafonova, I.I., 2000, \aap, submitted (astro-ph/0011513)
 

\reference{}
Lopez, S.,Reimers, D., Rauch, M., Sargent, W.L., Smette, A.	1999,	\apj,	513,	598

\reference{}
Lu, L., Savage, B.D., Tripp, T.M., \& Meyer, D. 1995, \apj, 447, 597

\reference{}
Lu, L., Sargent, W.L.W., \& Barlow, T.A. 1996, \apjs,  107, 475

\reference{}
Lu, L., Sargent, W.L.W., \& Barlow, T.A. 1998, \aj,  115, 55

\reference{}
Madau, P., 1992, \apj, 389, L1

\reference{}
Madau, P.,   Haardt, F., \& Rees, M.J. 1999, \apj, 514, 648

 
\reference{}
Molaro, P., Bonifacio, P., Centuri\'on, M., D'Odorico, S., Vladilo, G., Santin, P.,
Di Marcantonio, P., 2000, \apj, 541, 54

\reference{}
Nussbaumer, H., \& Storey, P.J. 1986, \aaps, 64, 545 (NS86)
 
\reference{}
Pei, Y.C., Fall, S.M., \& Bechtold, J. 1991, \apj, 378, 6
\reference{}

 
 
\reference{}
Pettini, M., Ellison, S.L., Steidel, C.C., Shapley, A.E., \& Bowen, D.V. 2000, \apj, 532, 65

\reference{}
 Pettini, M., King, D.L., Smith, L.J., \& Hunstead, R.W. 1997, \apj, 478, 536

\reference{}
Prochaska, J.X., Naumov, S.O.,  Carney, B.W., McWilliam, A., \& Wolfe, A.M.
2000, \aj, 120, 2513
 

\reference{}
Prochaska, J.X., \& Wolfe, A.M. 	1996,	 \apj, 470, 403

\reference{}
Prochaska, J.X., \& Wolfe, A.M. 	1997,	 \apj,	474,	140

\reference{}
Prochaska, J.X., \& Wolfe, A.M. 1999, \apjs, 121, 369


\reference{}
Savaglio, S.,  Panagia, N.,  Stiavelli, M., 2000, "Cosmic Evolution and Galaxy
Formation: Structure, Interactions, and Feedback", 
J. Franco, E. Terlevich, O. L\'opez-Cruz and I. Aretxaga, eds., A.S.P. Conf. Ser., in press,
(astro-ph/9912112)


\reference{}
Sembach, K.R., Howk, J.C., Ryans, R.S.I., \& Keenan, F.P. 2000, \apj, 528, 310

 
\reference{}
Sofia, U.J., \& Jenkins, E.B. 1998, \apj, 499, 951

\reference{}
Steidel, C.C., Pettini, M., \& Adelberger, K.L. 2000, \apj, in press (astro-ph/0008283)

\reference{}
Storrie-Lombardi, L .J.,  Irwin, M. J., \& McMahon, R. G. 	1996,	\mnras,	282,	1330

\reference{}
Verner, D.A., Ferland, G.J., Korista, K.T., Yakovlev, D.G. 1996,
\apj, 465, 487

\reference{}
Vladilo, G. 1998, \apj, 493, 583  (Paper I)

 

\reference{}
Vladilo, G., Bonifacio, P., Centuri\'on, M., Molaro, P. 2000, \apj, 543, 24

\reference{}
Viegas, S.M., 1995, \mnras, 276, 268

\reference{}
Wolfe, A.M., Lanzetta, K.M., Foltz, C.B., Chaffee, F.H., 1995, \apj, 454, 698

\reference{}
Wolfe, A.M., Prochaska, J.X., 2000, \apj, in press (astro-ph/0009081)

 
\end{references}
\end{document}